\documentclass[%
 reprint,
superscriptaddress,
 amsmath,amssymb,
 aps,
pra,
floatfix,
]{revtex4-2}

\usepackage{graphicx}
\usepackage{dcolumn}
\usepackage{bm}
\usepackage[]{xcolor}
\usepackage{subcaption}

\begin{document}


\title{Making the zeroth-order process fidelity independent
of state preparation and measurement errors}

\author{Yu-Hao Chen}
\email{d13222001@ntu.edu.tw}
\affiliation{Department of Physics, National Taiwan University, Taipei, Taiwan}%
\affiliation{Department of Artificial Intelligence, Chang Gung University, Taoyuan, Taiwan 
}

\author{Renata Wong}%
\thanks{Corresponding author}%
\email{renata.wong@cgu.edu.tw}
\affiliation{Department of Artificial Intelligence, Chang Gung University, Taoyuan, Taiwan 
}%
\affiliation{Department of Neurology, Chang Gung Memorial Hospital, Keelung, Taiwan 
}%
\affiliation{Artificial Intelligence Research Center, Chang Gung University, Taoyuan, Taiwan 
}

%
\author{Hsi-Sheng Goan}
\thanks{Corresponding author}%
\email{goan@phys.ntu.edu.tw}
\affiliation{Department of Physics, National Taiwan University, Taipei, Taiwan}
\affiliation{Center for Quantum Science and Engineering, National Taiwan University, Taipei, Taiwan}
\affiliation{Physics Division, National Center for Theoretical Sciences, Taipei, Taiwan}%


\date{\today}

\begin{abstract}

We demonstrate that the zero-fidelity, an approximation to the process fidelity, becomes robust to state preparation and measurement (SPAM) errors when combined with randomized benchmarking. However, as randomized benchmarking requires randomly choosing an increasingly large number of Clifford elements from the Clifford group when the qubit number increases, this combination is also limited to quantum systems with up to three qubits. To make the zero-fidelity independent of SPAM errors and, at the same time, applicable to multi-qubit systems, we employ a channel noise scaling method called zero noise extrapolation and show that this also makes the zero fidelity robust to SPAM errors. 

\end{abstract}

\maketitle


\section{\label{sec:level1}Introduction}

Evaluating the performance of physical quantum devices requires efficient and accurate characterization protocols. While process fidelity is a standard metric for this task \cite{Flammia}, its experimental implementation is highly resource-intensive . The protocol demands a considerable number of experimental settings, including frequent, shot-by-shot changes to initial states and measurement bases. Furthermore, because the mathematical formulation of process fidelity relies on Pauli operators, which are traceless and therefore not valid quantum states, experimentalists must prepare the eigenstates of these operators. For a single qubit, this requires preparing eight distinct initial states.

To mitigate these overheads, the zero-fidelity protocol \cite{Zero} was recently introduced as a zeroth-order approximation to the process fidelity. By utilizing a set of four symmetric, informationally complete (SIC) states \cite{SIC}, which are valid quantum states, the protocol significantly reduces the necessary experimental settings in terms of both state preparation and measurement bases. However, the practical utility of the zero-fidelity protocol is currently limited by its susceptibility to state preparation and measurement (SPAM) errors. For instance, initial states may suffer from over-rotation errors during preparation, and measurement outcomes may flip with a certain probability. Because characterizing SPAM errors experimentally is notoriously difficult, it is crucial for benchmarking protocols to be inherently robust against them. If left unmitigated, SPAM errors, which compound as the number of qubits increases, can severely distort fidelity evaluations, yielding results that do not accurately reflect the actual channel dynamics.

In this work, we demonstrate how to render the zero-fidelity protocol robust against SPAM errors. We first combine zero-fidelity with randomized benchmarking (RB) \cite{randomized3, randomized4, programmable}, establishing SPAM-independence for systems of up to three qubits. Because the rapid expansion of the Clifford group makes RB impractical to scale as the qubit count increases, we subsequently apply zero noise extrapolation (ZNE) \cite{GiurgicaTiron2020DigitalZN}. We show that this channel noise scaling technique successfully decouples SPAM errors from zero-fidelity estimations, making the protocol viable for larger multi-qubit architectures.

The primary objective of this paper is to demonstrate the robustness and independence of the zero-fidelity protocol against state preparation and measurement (SPAM) errors. We validated this resilience through comprehensive quantum error simulations, an approach necessitated by the fact that executing these protocols on current physical quantum processors remains prohibitive due to limited quantum hardware accessibility, availability, and quality. For this reason, we leave execution on quantum processors as future research.

The remainder of this paper is structured as follows. Section \ref{Sec2} introduces the zero-fidelity protocol and SIC-states. Section \ref{Sec3} details the integration of the zero-fidelity protocol with both randomized benchmarking and zero noise extrapolation. Section \ref{Sec4} presents our numerical experimental results, demonstrating the robustness of our combined protocols. Section \ref{Sec5} concludes the paper.

\subsection*{Contributions}

In this work, we make the following contributions: 

\begin{itemize}
\item We demonstrate that integrating the zero-fidelity protocol with randomized benchmarking successfully mitigates the impact of SPAM errors. However, the scaling limitations inherent to randomized benchmarking restrict the practical application of this combined approach to systems of up to three qubits.

\item To extend SPAM-error resilience to larger multi-qubit architectures, we introduce an alternative method that combines the zero-fidelity protocol with zero noise extrapolation.

\item Through numerical simulations, we verify that integrating zero-fidelity with either randomized benchmarking or zero noise extrapolation effectively neutralizes SPAM-induced distortions. Furthermore, these experiments confirm that the integration of these error mitigation techniques does not distort the zero-fidelity itself, preserving its utility as an accurate approximation of process fidelity.
 
\item Our findings establish that the zero-fidelity protocol can be practically deployed across multi-qubit systems, offering a resource-efficient benchmarking tool that maintains high accuracy despite the presence of SPAM errors. 
\end{itemize}

\section{Zero-fidelity as an approximation to process fidelity} \label{Sec2}

This section briefly introduces symmetric, informationally complete (SIC) states and the standard formulation of process fidelity. The zero-fidelity protocol serves as the zeroth-order approximation to process fidelity within a hierarchy of k-fidelities. While standard process fidelity relies on density matrices prepared in the eigenstates of Pauli operators, zero-fidelity utilizes a set of four SIC-states for state preparation.

The original zero-fidelity protocol successfully approximates this metric, albeit without accounting for SPAM errors. Here, we augment the zero-fidelity formulation to incorporate SPAM error robustness. To verify the accuracy of our augmentation procedure, we compare the obtained results with those in which SPAM errors were not considered.

\subsection{Symmetric, informationally complete states}

Positive operator-valued measures (POVMs) mathematically describe measurements performed on quantum systems. A POVM consists of $w$ positive semidefinite operators $\{E_{i}\}_{i=1}^{w}$ that span a Hilbert space $\mathcal{H}$ of dimension $d=2^{n}$, where $n$ represents the number of qubits. These operators must satisfy the completeness relation, summing to the identity: $\sum_{i=1}^{w}E_{i}=I$. For an $n$-qubit system, the tensor product of Pauli operators, $P_{i}\in\{I,X,Y,Z\}^{\otimes n}$, typically serves as the set of POVMs.

Alternatively, for a single qubit ($n=1$), a $d$-dimensional Hilbert space can be spanned by a minimal set of $d^{2}=4$ operators. These correspond to the four SIC-states. Unlike Pauli operators, which are mutually orthogonal, SIC-states possess uniform transitional probabilities of $|\langle\psi_{i}|\psi_{j}\rangle|^{2}=\frac{1}{d+1}$ for $i\neq j$. This geometric property ensures that any initial state projects onto the remaining three states with equal probability. This specific set of POVMs is known as a symmetric, informationally complete POVM (SIC-POVM). For a single qubit, the four SIC-states are explicitly defined as:
\begin{equation}
	\begin{split}
		|\psi\rangle_1 &= |0\rangle \\
		|\psi\rangle_2 &= \frac{1}{\sqrt{3}}|0\rangle + \sqrt{\frac{2}{3}}|1\rangle \\
		|\psi\rangle_3 &= \frac{1}{\sqrt{3}}|0\rangle + \sqrt{\frac{2}{3}}e^{i\frac{2\pi}{3}}|1\rangle \\
		|\psi\rangle_4 &= \frac{1}{\sqrt{3}}|0\rangle + \sqrt{\frac{2}{3}}e^{i\frac{4\pi}{3}}|1\rangle \\
		|\langle\psi_i|\psi_j\rangle|^2 &= \frac{1}{3}, \quad i \neq j.
	\end{split}
\end{equation}

On the Bloch sphere, these states map to the vertices of a regular tetrahedron and collectively constitute a quantum state 2-design \cite{Matteo2014ASI, 1523643}.

\subsection{Process fidelity and zero-fidelity}

Process fidelity quantifies the overlap between an ideal quantum channel $\mathcal{U}$ and an actual experimental channel $\Gamma$, effectively indicating how closely the executed quantum operation matches the theoretical ideal. This metric (after \cite{Zero, cycle}) is given by:
 \begin{equation}  \label{eq 10}
	F(\mathcal{U}, \Gamma) = \frac{1}{4^n}\sum^{4^n}_{i = 1}\mathbf{Tr}[\mathcal{U}(\sigma^{\dagger}_i)\Gamma(\sigma_i)],
\end{equation} 
where $\sigma_{i}$ represents unnormalized Pauli operators. This exact formulation, however, is experimentally inaccessible; Pauli operators are traceless and thus cannot represent valid physical quantum states prepared in a laboratory. To accommodate the requirement that inputs must be valid density matrices, process fidelity is rewritten as: 
\begin{equation}\label{eq 11}
F(\mathcal{U},\Gamma)=\frac{1}{4^{n}}\sum_{i,j=1}^{4^{n}}[B^{-1}]_{ij}\mathbf{Tr}[\mathcal{U}(\rho_{i})\Gamma(\rho_{j})],
\end{equation}
where $B_{ij}=\mathbf{Tr}[\rho_{i}^{\dagger}\rho_{j}]$ normalizes the equation and computes the overlap between states $\rho_{i}$ and $\rho_{j}$. If these density matrices are prepared in the eigenstates of Pauli operators, then the coefficient 
$$\begin{cases}
	B_{ij} = 2^n, & \mbox{if } i = j \\
	B_{ij} = 0, & \mbox{if } i \neq j.
\end{cases}$$  
By \cite{cycle}, Eq.~(\ref{eq 11}) will then reduce to 
\begin{equation} \label{eq 12}
	F(\mathcal{U}, \Gamma) = \frac{1}{4^n}\sum^{4^n}_{i,j = 1}2^{-n}\mathbf{Tr}[\mathcal{U}(\rho_i)\Gamma(\rho_i)].
\end{equation}
In accordance with Born's rule, the term $\mathbf{Tr}[\mathcal{U}(\rho_{i})\Gamma(\rho_{j})]$ represents the expectation value of the observable $\mathcal{U}(\rho_{i})$ evaluated against the final state $\Gamma(\rho_{j})$. 

While mathematically rigorous, this formulation remains practically challenging because the target observables $\mathcal{U}(\rho_{i})$ often exhibit entangled eigenstates. Entangled eigenstates cannot be factorized into individual qubit states and therefore require collective measurement of the entire system, a procedure that is highly complex or outright difficult on current hardware. Consequently, the metric had better be recast again using a set of strictly separable Pauli observables $W_{l}\in\{I,X,Y,Z\}^{\otimes n}$:
\begin{equation} \label{eq 13}
	F(\mathcal{U}, \Gamma) = \frac{1}{4^n}\sum^{4^n}_{i,l}C_{il}\mathbf{Tr}[\Gamma(\rho_i)W_l], 
\end{equation} 
where 
\begin{equation}
	C_{il} = \sum^{4^n}_{j = 1}[B^{-1}]_{ji}\mathbf{Tr}[\mathcal{U}(\rho_j)W_l], 
\end{equation} 
derived in detail in Appendix~\ref{appendix-a}.

Because this exact formulation requires an unscalable $16^{n}$ experimental settings ($4^{n}$ initial states paired with $4^{n}$ observables), researchers often sample a small subset of these settings via joint probability distributions to reduce overhead \cite{Flammia}. To facilitate this sampling without suffering the inefficiencies of frequently reconfiguring initial states and measurement bases, one can deploy the zero-fidelity approximation \cite{Zero}:
\begin{equation}\label{eq 15}
F_{0}(\mathcal{U},\Gamma)=\frac{1}{4^{n}}\sum_{i,j}^{4^{n}}\mathbf{Tr}[\mathcal{U}(\rho_{i})W_{j}]\mathbf{Tr}[\Gamma(\rho_{i})W_{j}],
\end{equation}
where the input states $\rho_{i}$ are prepared in the four SIC-states rather than Pauli eigenstates. While highly efficient, this fundamental formulation of zero-fidelity does not natively account for state preparation and measurement errors. The subsequent sections detail how we integrate mitigating techniques into this equation to fully decouple SPAM errors from the estimation.

\section{Zero-fidelity randomized benchmarking and channel noise scaling} \label{Sec3}

This section details the theoretical framework for decoupling SPAM errors from zero-fidelity estimation. We first review standard randomized benchmarking (RB) \cite{EmersonRBCYCLE} and adapt its formalism for the zero-fidelity protocol. Subsequently, to circumvent the restrictive scaling limits inherent to RB, we introduce a channel noise scaling technique known as zero noise extrapolation (ZNE).

\subsection{Randomized benchmarking and process fidelity} \label{3.1}

Based on the randomized benchmarking protocol \cite{EmersonRBCYCLE}, the process fidelity can be made robust to SPAM errors by applying quantum operations of the noisy and noiseless channels for Cllifford length $m$ times and gradually increasing the value of $m$. Therefore the process fidelity equation can be written in the following form
\begin{equation} \label{eq 16}
    F^m(\mathcal{U}, \Gamma) = \frac{1}{4^n}\sum^{4^n}_{i = 1}\mathbf{Tr}[\mathcal{U}^m(\sigma^{\dagger}_i)\Gamma^m(\sigma_i)],
\end{equation} 
where $\sigma_i$ is the tensor product over single-qubit Pauli operators. However, as mentioned in Section \ref{Sec2}, Eq.~(\ref{eq 16}) cannot be estimated experimentally because Pauli operators are traceless. Instead, one needs to prepare the eigenstates of Pauli operators and avoid using observables $\Lambda(\rho_i)$ with entangled eigenstates. For the above reasons, Eq.~(\ref{eq 16}) is further rewritten as  
\begin{equation}
    F^m(\mathcal{U}, \Gamma) = \frac{1}{4^n}\sum^{4^n}_{i,j = 1}2^{-n}\mathbf{Tr}[\mathcal{U}^m(\rho_i)W_j]\mathbf{Tr}[\Gamma^m(\rho_i)W_j].
\end{equation}

Based on the reasoning in both randomized benchmarking and cycle benchmarking \cite{EmersonRBCYCLE, cycle}, one can also decouple SPAM errors from the zero-fidelity. In such a scenario, Eq.~(\ref{eq 15}) becomes 
\begin{equation}
    F_0^m(\mathcal{U}, \Gamma) = \frac{1}{4^n}\sum^{4^n}_{i,j = 1}\mathbf{Tr}[\mathcal{U}^m(\rho_i)W_j]\mathbf{Tr}[\Gamma^m(\rho_i)W_j].
\end{equation} \label{eq 17}

We now demonstrate how to combine process fidelity and randomized benchmarking for different Clifford lengths $m$. 
A quantum state $\rho$ that evolves through an ideal channel $\mathcal{U}$ can be described as \begin{equation}
    \rho \rightarrow \mathcal{U}(\rho) = U\rho U^{\dagger}, 
\end{equation}  where $U$ is a unitary operator.
Similarly, a quantum state that evolves through a noisy channel can be described as \begin{equation} \label{eq 19}
    \rho \rightarrow \tilde{\mathcal{U}}(\rho) = \tilde{U}\rho \tilde{U}^{\dagger}.
\end{equation} Assume that the noisy channel $\tilde{\mathcal{U}}$ can be decomposed into $\Lambda\mathcal{U}$ so that Eq.~(\ref{eq 19}) becomes \begin{equation}
     \rho \rightarrow \tilde{\mathcal{U}}(\rho) = \Lambda(U\rho U^{\dagger}).
\end{equation} Therefore, from the process fidelity we have \begin{equation}
    F(\mathcal{U}, \tilde{\mathcal{U}}) = \frac{1}{4^n}\sum^{4^n}_{i = 1}\mathbf{Tr}[\rho_i U^{\dagger}  \Lambda(U\rho_i U^{\dagger}) U].
\end{equation} Now, we consider ideal channels and noisy channels applied to a quantum state $m$ times, which we define as \begin{equation}
    \rho'_i = \tilde{\mathcal{C}}^m\rho_i = \Lambda\mathcal{U}^{-1}\Lambda\mathcal{U}_m...\Lambda\mathcal{U}_2\Lambda\mathcal{U}_1\rho_i
\end{equation} and \begin{equation}
    \rho'_i = \mathcal{C}^m\rho_i = \mathcal{U}^{-1}\mathcal{U}_m...\mathcal{U}_2\mathcal{U}_1\rho_i,
\end{equation} where $ \mathcal{U}^{-1} = (\mathcal{U}_m...\mathcal{U}_2\mathcal{U}_1)^{-1}.$ 

Using the superoperator representation $\Lambda \rightarrow \hat{\Lambda} = \sum_{k}A^{*}_k \otimes A_k$ and $\mathcal{U} \rightarrow \hat{U} =U^{*} \otimes U$, where $A_k$ and $U$ are operator and unitary operator, respectively  and * denotes complex conjugation, we obtain the process fidelity as \begin{align*}
    F(\mathcal{C}^m, \tilde{\mathcal{C}}^m) &= \frac{1}{4^n}\sum^{4^n}_{i = 1}\mathbf{Tr}[\mathcal{C}^m(\rho_i)\tilde{\mathcal{C}}^m(\rho_i)] \\ &= 
    \frac{1}{4^n}\sum^{4^n}_{i = 1}\mathbf{Tr}[\rho_i(\hat{U}^{-1}_{m}\hat{\Lambda}_{m}\hat{U}_m\cdots \hat{U}^{-1}_{1}\hat{\Lambda}_{1}\hat{U}_1)\rho_i]. 
\end{align*} From $\mathbb{E}_{U}(F_g) = \mathbf{Tr}\big[\rho(\int_{U(D)} dU \hat{U} \hat{\Lambda} \hat{U}^{-1})\rho\big]$ (the gate fidelity over all unitary operators by Haar measurement \cite{EmersonRBCYCLE}), we obtain the formula for averaging over the Haar measure for the unitary operators 
\begin{align*}
    F_{\text{avg}}(\mathcal{C}^m, \tilde{\mathcal{C}}^m) &= \mathbb{E}_{U}(F(\mathcal{C}^m,\tilde{\mathcal{C}}^m)) \\ &= \int_{U(D^{\otimes n})} (\Pi^m_{j = 1 }dU_j)(F(\mathcal{C}^m,\tilde{\mathcal{C}}^m)) \\ &=
    \frac{1}{4^n}\sum^{4^n}_{i = 1}\mathbf{Tr}(\rho_i\Lambda[\Pi^m_{j = 1}\hat{\Lambda}^{avg}_{j}]\rho_i),
\end{align*} where $dU_j$ denotes the Haar measure and \begin{equation*}
    \hat{\Lambda}^{\text{avg}}_{j} \equiv \mathbb{E}_{U_j}(\hat{\Lambda}_j) \equiv \int_{U(D)} dU \hat{U}^{-1}\hat{\Lambda}_{j}\hat{U}.
\end{equation*} As given in \cite{AVE}, \begin{equation*}
    \hat{\Lambda}^{\text{avg}}_{j}\rho = p_j\rho+(1-p_j)\frac{\mathbb{I}}{D}
\end{equation*} with \begin{equation}\label{pj}
    p_j = \frac{\mathbf{Tr}(\hat{\Lambda}_j)-1}{D^2 - 1}.
\end{equation} With the above, we can finally obtain the relation \begin{equation}
\begin{split}
    F^m(\mathcal{U}, \tilde{\mathcal{U}}) &= \frac{1}{4^n}\sum^{4^n}_{i = 1}\mathbf{Tr}[\mathcal{U}^m(\rho_i)\tilde{\mathcal{U}}^m(\rho_i)] \\ &= A_0p^m + B_0, \label{eq 35}
\end{split}
\end{equation} where the coefficients $A_0$ and $B_0$ absorb SPAM errors and the strength of depolarizing error $p$ can be considered as the parameter for the average gate fidelity $F_{\text{avg}} = p+\frac{1-p}{2^n}$. Here, we assume the same strength $p_j$ as in Eq.~(\ref{pj}), gate-independence and time-independence \cite{EmersonRBCYCLE, RBassumption}. See Appendix \ref{Appen C} for an analysis of the physical meaning of the parameter $p$.

Now, from \cite{Flammia} one can also estimate the process fidelity by the following equation  \begin{equation} \label{eq 36}
    F^m(\mathcal{U}, \tilde{\mathcal{U}}) = \frac{1}{4^n}\sum^{4^n}_{i, j =1}2^{-n}\mathbf{Tr}[\mathcal{U}^m(\rho_i)W_j]\mathbf{Tr}[\tilde{\mathcal{U}}^m(\rho_i)W_j],
\end{equation} where the quantum states $\rho_i$ are prepared in the eigenstates of Pauli operators.

\subsection{Zero-fidelity with randomized benchmarking}

To render the zero-fidelity protocol robust against SPAM errors while preserving its reduced experimental overhead, we integrate it directly with the RB framework. From Eq.~(\ref{eq 36}) we can write down the equation for the zero-fidelity as 
\begin{equation} \label{eq 37}
	F^m(\mathcal{U}, \tilde{\mathcal{U}}) = \frac{1}{4^n}\sum^{4^n}_{i, j =1}\mathbf{Tr}[\mathcal{U}^m(\rho_i)W_j]\mathbf{Tr}[\tilde{\mathcal{U}}^m(\rho_i) W_j],
\end{equation} 
where the quantum states $\rho_i$ are prepared by the SIC-states. As derived in Appendix B, this modified zero-fidelity formulation in Eq.~(\ref{eq 37}) inherits the SPAM-independence of standard RB.

In practical circuit implementations, we generate these RB sequences utilizing the Gottesman-Knill theorem \cite{RBCIRCUIT} to efficiently compute the inverse of the initial $m$ Clifford operators. To optimize classical memory and facilitate execution on physical backend models (e.g., Qiskit \cite{Qiskit}), we compile these Clifford elements into a restricted base gate set comprising Hadamard, Phase, CNOT, and Pauli gates. This structural decomposition exploits the fact that the expansive $n$-qubit Clifford group is simply the semidirect product of the much smaller symplectic group and the Pauli group \cite{Threerb}, vastly reducing overhead.

\subsection{Channel noise scaling} \label{C}

While integrating zero-fidelity with RB successfully isolates SPAM errors, the exponential growth of the Clifford group practically limits this approach to systems of up to three qubits. To extend SPAM-robust zero-fidelity estimation to larger architectures, we bypass RB entirely and employ zero noise extrapolation (ZNE) \cite{GiurgicaTiron2020DigitalZN}. ZNE systematically amplifies the noise of a target quantum circuit to extract a decay rate independent of the initial and final states. 

Specifically, we utilize global unitary folding \cite{GiurgicaTiron2020DigitalZN}, wherein the target channel $\mathcal{U}$ is repeatedly applied as $(\mathcal{U}\mathcal{U}^{\dagger})^{m}=\mathcal{I}^{m}$ for $m \ge 1$. This effectively increases the depth, and consequently, the accumulated error, of the target gate block without modifying the state preparation or measurement layers. For depolarizing noise channels, repeatedly applying these identity circuits (identity folding) yields an exponential decay curve identical in form to that observed in randomized benchmarking. By fitting the zero-fidelity estimations as a function of the folding depth $m$, we can extract a fidelity value that is strictly intrinsic to the channel and entirely independent of SPAM errors. Crucially, because ZNE bypasses the need to sample from the Clifford group, it dramatically reduces the memory footprint and the sheer volume of required circuits, rendering the protocol highly scalable for multi-qubit systems.

\begin{figure}
    \centering
    \includegraphics[width=9cm]{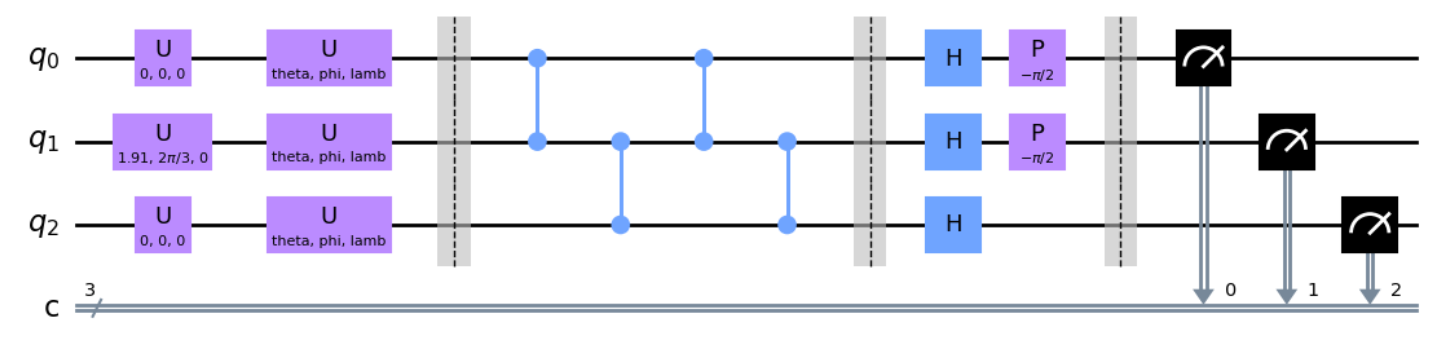}
    \caption{\footnotesize{An example three-qubit circuit. The first U gate on each qubit is for state preparation, where we rotate $|0\rangle$ to one of the SIC-states. The second U gate on each qubit is state preparation error. The actual target channel consists of a block of CZ gates. The gates behind the CZ gates are for performing measurement basis changes. We also add readout error on each measurement gate. The ideal target channel contains no SPAM or gate errors.}}
    \label{fig CZ}
\end{figure}

As an example, let's assume that we want to calculate the zero-fidelity of a channel $\mathcal{U}$ whose circuit looks like the one given in Fig.~\ref{fig CZ}. In Fig.~\ref{fig CZ}, the first U gate on each qubit is for state preparation, where we rotate $|0\rangle$ to one of the SIC-states. The second U gate on each qubit is a state preparation error. Then follows the actual target channel, which contains a block of CZ gates. The gates behind the CZ gate block are for performing measurement basis changes. We also add readout error on each measurement gate. 

We can apply the same channel $\mathcal{U}$ a number of $m$ times, i.e. $(\mathcal{UU^{\dagger}})^m = \mathcal{I}^m$, where $m \geq 1$. 
This method is called global folding. 

By applying global folding to the circuit in Fig.~\ref{fig CZ}, we effectively extend only the CZ block of the circuit $m$ times. The gates before the block as well as the gates after the block remain intact. So, for $m=1$, the circuit will consist of a CZ block containing a total of 8 CZ gates, while for $m=2$, the circuit will have a block of 16 CZ gates, etc. This method produces a total of $m+1$ circuits for a given $m$.

In the ideal case, repeatedly applying the same circuit will not affect the final result. However, the error will increase with the number of channels applied in actual experiments. This indicates that the zero-fidelity can be seen as a function of the number of circuits. With this, one can then also obtain a decay rate by fitting the results of the zero-fidelity against the number of circuits.

Identity folding has been used for mitigating errors in quantum computation for instance in \cite{GiurgicaTiron2020DigitalZN, Mitigation1, Mitigation2}. It has been shown experimentally that if the noise is depolarizing noise (and this holds also in the case of gate-dependence) the results can be fitted with the same exponential decay as in randomized benchmarking \cite{GiurgicaTiron2020DigitalZN}. Using this method, we don't have to select Clifford gates randomly from the Clifford group in order to calculate the average gate fidelity. Instead, we can obtain a fidelity value from the zero-fidelity decay rate, and this value will be independent of SPAM errors. Since we don't need to choose from the Clifford gates, we do not therefore need to deal with the high number of Clifford elements in the Clifford group. This implies the ability to obtain the zero-fidelity for systems of more than three qubits as well.

\section{Numerical results} \label{Sec4}

To validate the robustness of our proposed protocols, we simulated zero-fidelity estimations across systems of varying qubit scales using Qiskit's noisy backend models. In all simulations, gate errors were modeled as depolarizing noise applied to the CZ gates, mathematically expressed as:
\begin{equation}E(\rho)=(1-\lambda)\rho+\lambda \mathbf{Tr}[\rho]\frac{1}{2^{n}} ,
\end{equation}
where $\lambda$ parameterizes the gate error rate (fixed to $\lambda=0.01$ unless otherwise specified). To simulate SPAM errors, we introduced rotation errors sampled from a Gaussian distribution (mean $\mu=0$, standard deviation $\sigma=\sqrt{5}$) during state preparation, and explicit readout errors during measurement. Each experimental configuration was executed with 1024 shots. The following subsections detail the unmitigated baseline zero-fidelity (Section~\ref{tqzf}), followed by the results of applying randomized benchmarking (Section~\ref{zfrb}) and zero noise extrapolation (Section~\ref{zfzne}) to mitigate these SPAM errors.

\subsection{Baseline zero-fidelity without SPAM error mitigation}\label{tqzf}

\begin{widetext}
\begin{figure}[htbp]
    \centering
    \captionsetup[subfigure]{position=top, singlelinecheck=false,skip=0pt}%
    \begin{subfigure}[t]{0.31\textwidth} 
        \centering
        \includegraphics[width=\textwidth]{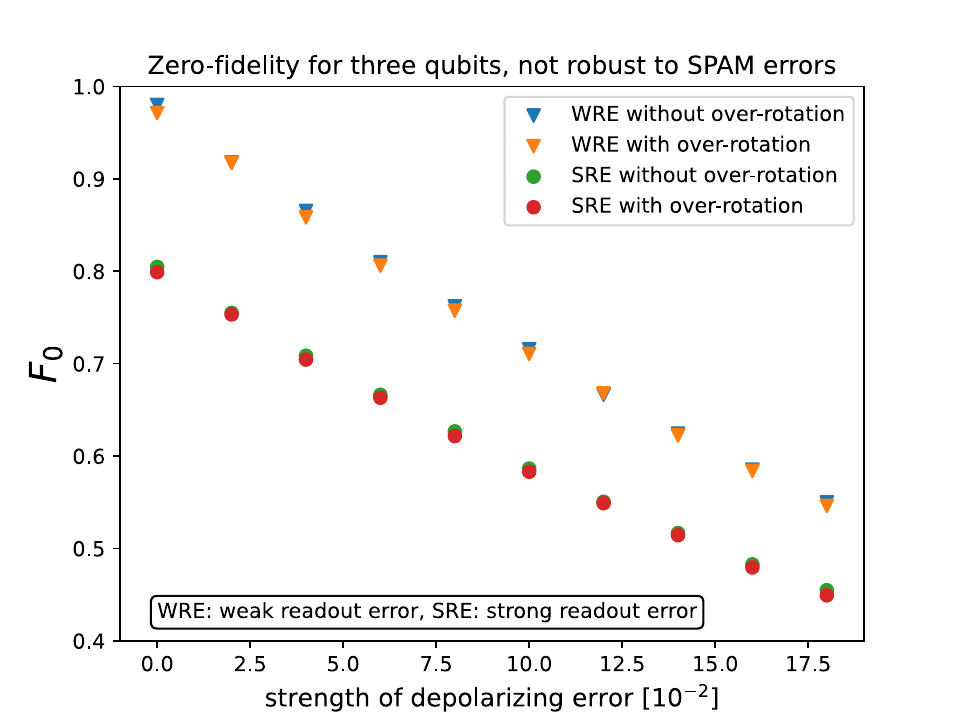} 
        \caption{Strong vs weak readout errors. Varying depolarizing error. }
        \label{fig:subfig1}
    \end{subfigure}%
    \begin{subfigure}[t]{0.31\textwidth}
        \centering
        \includegraphics[width=\textwidth]{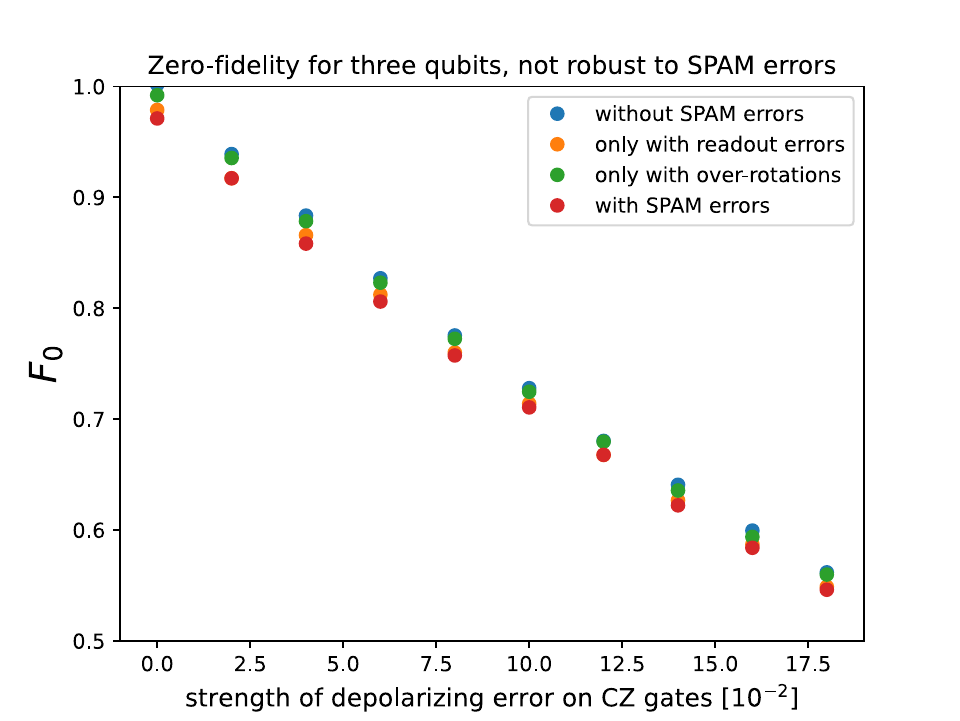}
        \caption{Varying depolarizing error strength, and weak readout error.}
        \label{fig 1}
    \end{subfigure}%
    \begin{subfigure}[t]{0.31\textwidth}
        \centering
        \includegraphics[width=\textwidth]{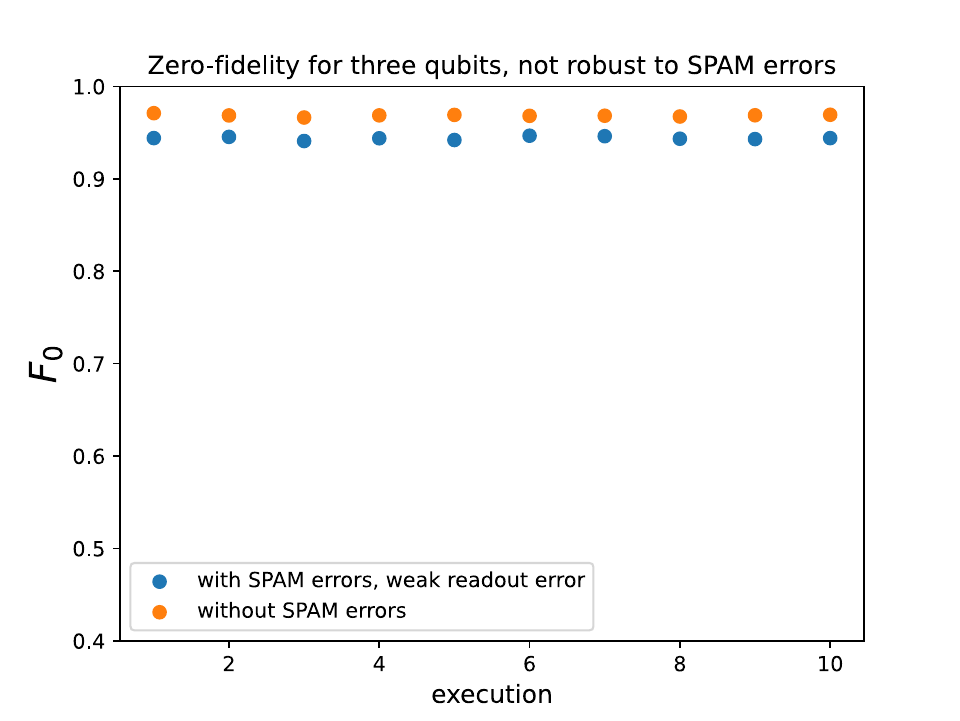}
        \caption{Fidelity at fixed gate error $\lambda=0.01$. }
        \label{fig:CZ_SPAM_fixed_gate_error}
    \end{subfigure}%
    \caption{Zero-fidelity for three-qubit circuit as in Fig.~\ref{fig CZ}. }
    \label{fig:zero-fidelity}
\end{figure}
\end{widetext}

To establish a baseline, we first evaluated the standard zero-fidelity protocol on a three-qubit target circuit (Fig. 1) without applying any error mitigation techniques. The target channel consisted of a block of CZ gates. To capture the full spectrum of state preparation and measurement configurations, each execution comprised 4096 distinct circuit combinations, reflecting the $4^3$ possible state preparations and $4^3$ possible measurement bases required for three qubits. We evaluated the circuit under both weak and strong readout errors, defined by the following transition matrices, respectively:
\begin{equation}\label{read}
E_{\text{strong}} = \begin{pmatrix}0.97&0.03\\0.05&0.95\end{pmatrix}, \end{equation}
\begin{equation}\label{read_s}
	E_{\text{weak}} = \begin{pmatrix}0.997&0.003\\0.005&0.995\end{pmatrix}.\end{equation}
As illustrated in Fig.~\ref{fig:subfig1} the unmitigated zero-fidelity $F_0$ predictably degrades as both the depolarizing error rate and the readout error strength increase. While over-rotations during state preparation lower the overall fidelity, they are less detrimental than the combined effects of readout and depolarizing errors. Fig.~\ref{fig 1} isolates these specific error modalities as a function of increasing depolarizing noise. Executions conducted in an ideal, SPAM-free environment consistently yielded the highest fidelities. Introducing isolated SPAM elements (either state preparation over-rotations alone or weak readout errors alone) depressed the fidelity, with the complete SPAM error model yielding the lowest estimates. Finally, Fig.~\ref{fig:CZ_SPAM_fixed_gate_error} demonstrates this discrepancy at a fixed gate error rate of $\lambda=0.01$. The average zero-fidelity evaluated without SPAM errors was 0.968, whereas the introduction of full SPAM errors artificially suppressed the estimated fidelity to 0.943. This confirms that unmitigated zero-fidelity is highly sensitive to SPAM errors, necessitating the scaling protocols discussed in the subsequent sections.

\subsection{Zero-fidelity with randomized benchmarking}\label{zfrb}

To mitigate the SPAM errors observed in the baseline estimations, we integrated the zero-fidelity protocol with randomized benchmarking (RB). Experimentally, this was achieved by applying sequences of $m+1$ operators uniformly sampled from the Clifford group. For each sequence, the initial $m$ operators were chosen randomly, while the final $(m+1)$-th operator was computed as the exact inverse of the preceding sequence to ensure an ideal net operation of identity.

For a given sequence length $m$, we generated multiple random sequences and applied them to the initial SIC-states. We then measured the expectation values of the Pauli observables to calculate the sequence-specific fidelity. These results were subsequently averaged over all random sequences to obtain the averaged sequence fidelity, $F_{\text{seq}}^{m}$. Finally, these values were fitted to the exponential decay model $F_{\text{seq}}^{m} = A_0 p^m + B_0$.

\begin{figure}
    \centering
    \includegraphics[width=0.5\textwidth]{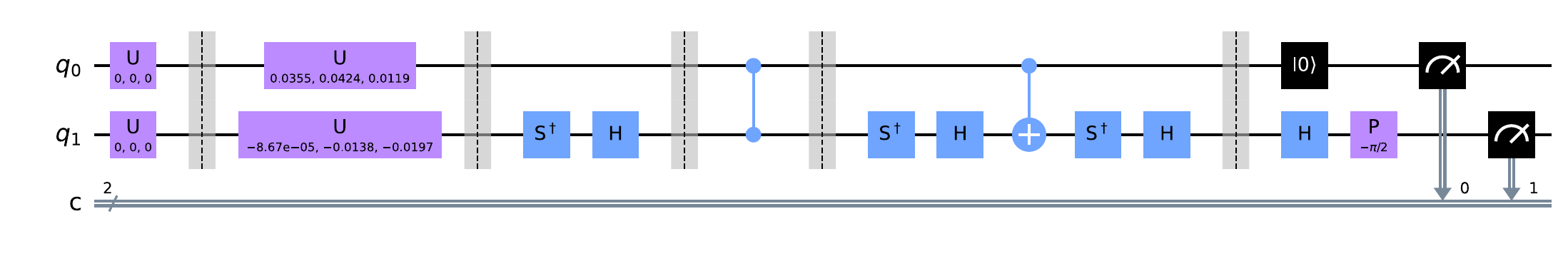}
    \caption{\footnotesize{A simple example a two-qubit circuit for zero-fidelity combined with interleaved randomized benchmarking. The first block of U gates is for preparation of SIC states, the second block of U gates is the state preparation error. Then, we have a Clifford gate represented by two operations, followed by the interleaved CZ channel, and the inverse of the Clifford gate. The gates directly before the measurements are used for basis change. For regular randomized benchmarking, we don't consider the interleaved CZ channel. 
    }}
    \label{fig:rb-circuit}
\end{figure}

The concrete steps utilized in the integration were as follows: 

\begin{enumerate}
\item Apply a sequence of $m+1$ operators chosen uniformly from the Clifford group. The first $m$ operators can be chosen randomly, however the choice of the $(m+1)$-th operator depends on the first $m$ operators because the $(m+1)$-th operator is the inverse of all the other $m$ operators taken together. Formally, let $C$ be an operator and let $S = C_{m+1}\circ C_m \circ...C_3\circ C_2\circ C_1$, where $\circ$ denotes composition of operators. For simplicity, we write $C_{m+1}\circ C_m \circ...C_3\circ C_2\circ C_1$ as $\bigcirc^{m+1}_{j = 1}C_j$. Then we require that $C_{m+1} = (C_m \circ...C_3\circ C_2\circ C_1)^{-1}$. If the operators are ideal (i.e., no noise is present), the sequence applied onto the quantum state is $S(\rho) = \rho.$ For a noisy quantum operator $\tilde{C}=\Lambda \circ C$ and a sequence $\tilde{S}$, we can describe them as $\tilde{S}^m = \bigcirc^{m+1}_{j = 1}(\Lambda\circ C_j). $ Here, we generate more than one sequence $\tilde{S}^{m}_{l} = \bigcirc^{m+1}_{j = 1}(\Lambda\circ C_{l_j})$ and then apply all of the generated sequences on the initial states. An example circuit for two-qubit systems is given in Fig.~\ref{fig:rb-circuit}.

\item For each sequence, calculate the expectation values of Pauli operators $W_j$ with respect to the final states  $S^{m}_{l}(\rho_i)$ and $\tilde{S}^{m}_{l}(\rho_i)$, respectively, where $\rho_i$ are the initial states, i.e. the SIC-states. After that, calculate the overlap between $S^{m}_{l}(\rho_i)$ and $\tilde{S}^{m}_{l}(\rho_i)$ as follows: \begin{equation} \label{4.37}
    \frac{1}{4^n}\sum^{4^n}_{i,j}\mathbf{Tr}[S^{m}_{l}(\rho_i)W_j]\mathbf{Tr}[\tilde{S}^{m}_{l}(\rho_i)W_j].
\end{equation}

\item Estimate Eq.~(\ref{4.37}) for different sequence $l$ and different Clifford length $m$ and then average the random sequence to find the averaged sequence fidelity: \begin{equation} \label{4.38}
    F^m_{\text{seq}}(\mathcal{S}^m, \tilde{\mathcal{S}}^m) = \frac{1}{4^n}\sum^{4^n}_{i,j}\mathbf{Tr}[\mathcal{S}^m(\rho_i)W_j]\mathbf{Tr}[\tilde{\mathcal{S}}^m(\rho_i)W_j],
\end{equation} where $$\mathcal{S}^m = \frac{1}{L}\sum^{L}_{l = 1}S^{m}_{l}.$$

\item Use the results estimated using Eq.~(\ref{4.38}) to fit the model mentioned above: \begin{equation*}
    F^m_{\text{seq}}(\mathcal{S}^m, \tilde{\mathcal{S}}^m) = A_0p^m+B_0.
\end{equation*}

\item With the result from Step 4, plot the exponential decay curve of the process fidelity and Clifford length $m$.
\end{enumerate}

\begin{widetext}
	\begin{figure}[htbp]
		\centering
		\captionsetup[subfigure]{position=top, singlelinecheck=false,skip=0pt}%
		\begin{subfigure}[t]{0.32\textwidth} 
			\centering
			\includegraphics[width=\textwidth]{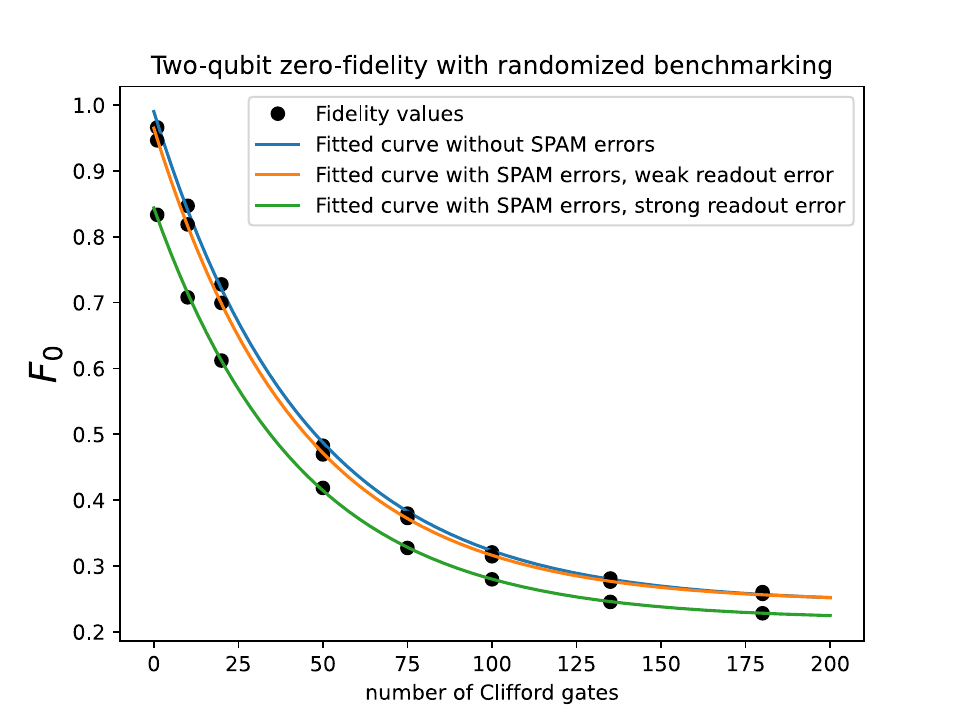} 
			\caption{Two-qubit system.}
			\label{fig 3}
		\end{subfigure}%
		\begin{subfigure}[t]{0.32\textwidth}
			\centering
			\includegraphics[width=\textwidth]{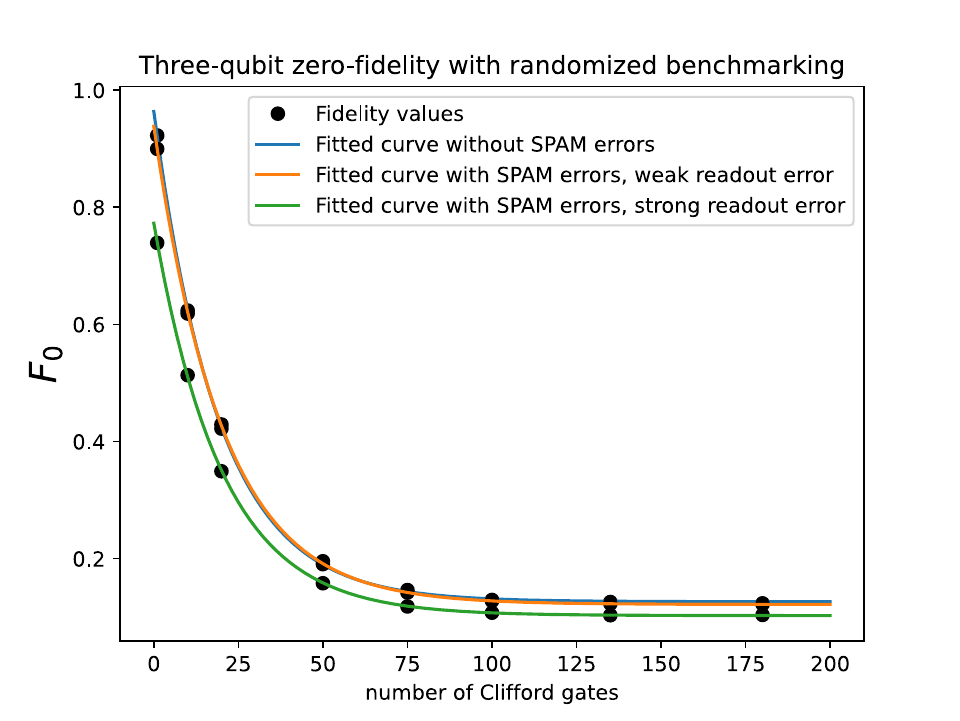}
			\caption{Three-qubit system.}
			\label{Three qubit RB}
		\end{subfigure}%
		\begin{subfigure}[t]{0.32\textwidth}
			\centering
			\includegraphics[width=\textwidth]{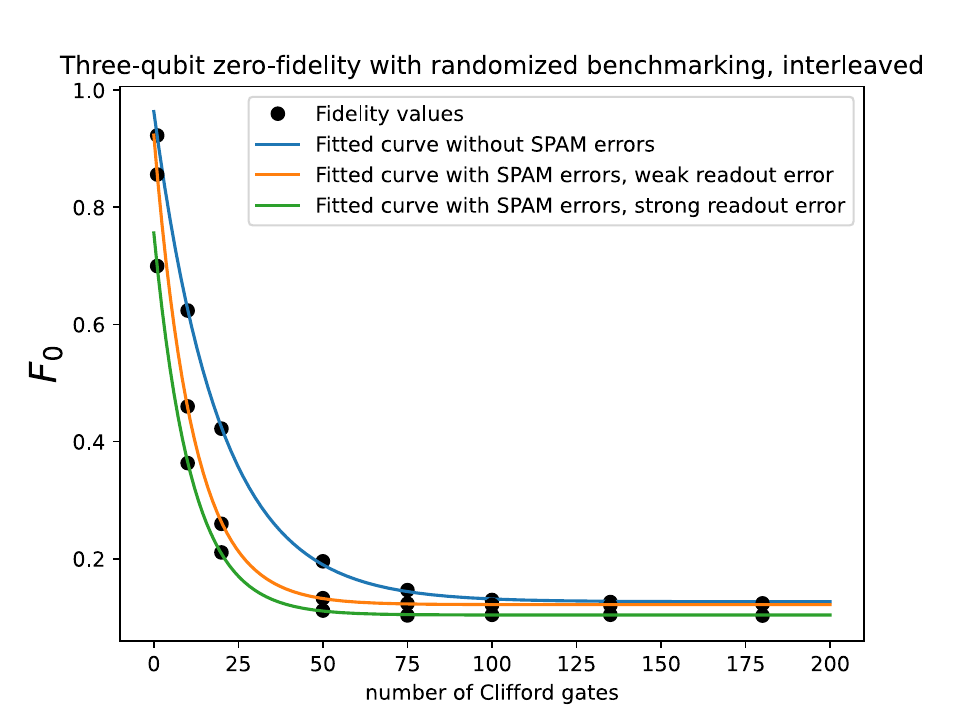}
			\caption{Three-qubit system. Only circuits with SPAM errors were interleaved.}
			\label{Three qubit RB_IRB weak SPAM}
		\end{subfigure}%
		\caption{Zero-fidelity with randomized benchmarking. }
		\label{fig:zero-fidelity-rb}
	\end{figure}
\end{widetext}

We executed this protocol on both two-qubit and three-qubit configurations. The numerical results are shown in Fig.~\ref{fig:zero-fidelity-rb}. All fidelity values were obtained under presence of depolarizing error. As depicted in Figs~\ref{fig 3} and ~\ref{Three qubit RB}, the zero-fidelity values at low sequence lengths are demonstrably degraded by the presence of SPAM errors. However, as the sequence length $m$ increases, the effect of these errors is exponentially suppressed, and the fidelity curves converge. Crucially, the statistical analysis (Tables I and II) confirms that the decay rate $p$ remains invariant regardless of the presence or strength of SPAM errors. For instance, in the two-qubit system, the decay rate remained perfectly constant at $p = 0.977$ across all error models. This behavior was mirrored identically in the three-qubit system, confirming that combining zero-fidelity with RB successfully isolates the intrinsic channel fidelity from state preparation and measurement imperfections.

To further validate this approach for specific gate characterization, we performed interleaved randomized benchmarking \cite{IRB, 24759412} on a three-qubit system. In this protocol, the target CZ channel (from Fig.~\ref{fig CZ}) was interleaved between the random Clifford gates. As shown in Fig.~\ref{Three qubit RB_IRB weak SPAM}, while the inclusion of additional interleaved gates inherently lowers the overall fidelity curves compared to standard RB, all curves (with and without SPAM errors) ultimately converge at large sequence lengths. By extracting the decay rate from the interleaved curves, we obtained an isolated zero-fidelity of $F_0 \simeq 0.967$ for the target CZ channel. This value closely aligns with the baseline SPAM-free fidelity observed in Fig.~\ref{fig:CZ_SPAM_fixed_gate_error}. From the above results one can observe that even when the strength of the readout error is increased, the same zero-fidelities of the target circuit are obtained. The curves with SPAM errors are lower than the curve without SPAM errors because we have included some extra gates for interleaved randomized benchmarking. However, they all converge to the same values when the number of Clifford gates is large. Therefore, the results demonstrate that by using randomized benchmarking with zero-fidelity we can calculate the zero-fidelity without the effects caused by SPAM errors.

\begin{table}[b]
\caption{
Statistics for the zero-fidelity with randomized benchmarking on two-qubit systems.
}
\begin{ruledtabular}\label{tab:stats}
\begin{tabular}{lrrr}
\textrm{Curve\footnote{WRE: weak readout error. SRE: strong readout error}}&
\textrm{Decay rate $p$}&
\textrm{Avg. error $r$\footnote{Avg. error per Clifford gate}}&
\textrm{$F_{\textbf{ave}}$}\\
\colrule
w/o SPAM err. & 0.977 & $1.71 \cdot 10^{-2}$ & 0.983\\
SPAM err., WRE & 0.977 & $1.71 \cdot 10^{-2}$ & 0.983\\
SPAM err., SRE & 0.977 & $1.71 \cdot 10^{-2}$ & 0.983\\
\end{tabular}
\end{ruledtabular}
\end{table}

\begin{table}[b]
	\caption{
		Statistics for the zero-fidelity with randomized benchmarking on three-qubit systems.
	}
	\begin{ruledtabular}\label{tab:stats2}
		\begin{tabular}{lrrr}
			\textrm{Curve\footnote{WRE: weak readout error. SRE: strong readout error}}&
			\textrm{Decay rate $p$}&
			\textrm{Avg. error $r$\footnote{Avg. error per Clifford gate}}&
			\textrm{$F_{\textbf{ave}}$}\\
			\colrule
			w/o SPAM err. & 0.952 & $4.20\cdot10^{-2}$ & 0.958\\
			SPAM err., WRE & 0.952 & $4.20\cdot10^{-2}$ & 0.958\\
			SPAM err., SRE & 0.952 & $4.23 \cdot 10^{-2}$ & 0.958\\
		\end{tabular}
	\end{ruledtabular}
\end{table}

\subsection{Zero-fidelity with zero noise extrapolation}\label{zfzne}

To demonstrate the scalability of our approach for systems exceeding three qubits, we evaluated the zero-fidelity using ZNE. This was realized via global unitary folding, which repeatedly applies identity layers to scale the channel noise without altering state preparation or measurement operations. Simulations were conducted on quantum systems of three, four, and five qubits. As with the previous experiments, the depolarizing gate error was fixed at $\lambda=0.01$. Because this noise is applied discretely to each CZ gate in the Qiskit model, the total accumulated depolarizing strength $p$ is a function of the gate count. For example, the theoretical depolarizing effect for four CZ gates is $0.99^{4}\approx0.961$. In our physical simulations, running 1024 shots introduces a minor projective sampling error, yielding an empirical baseline fidelity of 0.968. Therefore, the decay parameter $p$ extracted from the ZNE exponential fit serves as a precise indicator of the total depolarizing strength across the specific circuit volume.The effectiveness of ZNE is illustrated in Fig.~\ref{fig:zero-fidelity-zne}, which displays the fitted fidelity curves for the three-, four-, and five-qubit systems under identity folding. Consistent with our baseline findings, the raw fidelity values obtained under SPAM errors (both weak and strong readout models) are visibly depressed compared to the SPAM-free executions. However, fitting these decaying curves reveals that the decay rate $p$ remains rigorously invariant across all error conditions.

\begin{widetext}
\begin{figure}[htbp]
    \centering
    \captionsetup[subfigure]{position=top, singlelinecheck=false,skip=0pt}%
    \begin{subfigure}[t]{0.32\textwidth} 
        \centering
        \includegraphics[width=\textwidth]{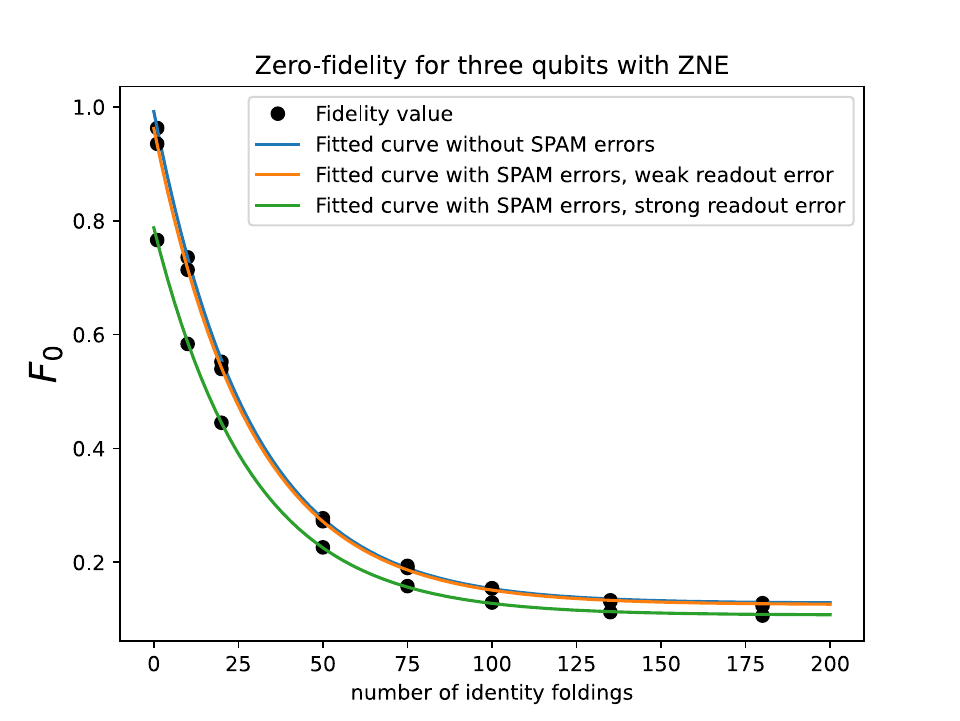} 
        \caption{$p \simeq 0.966$ for all curves.}
        \label{Zero_fidelity_channels}
    \end{subfigure}%
    \begin{subfigure}[t]{0.32\textwidth}
        \centering
        \includegraphics[width=\textwidth]{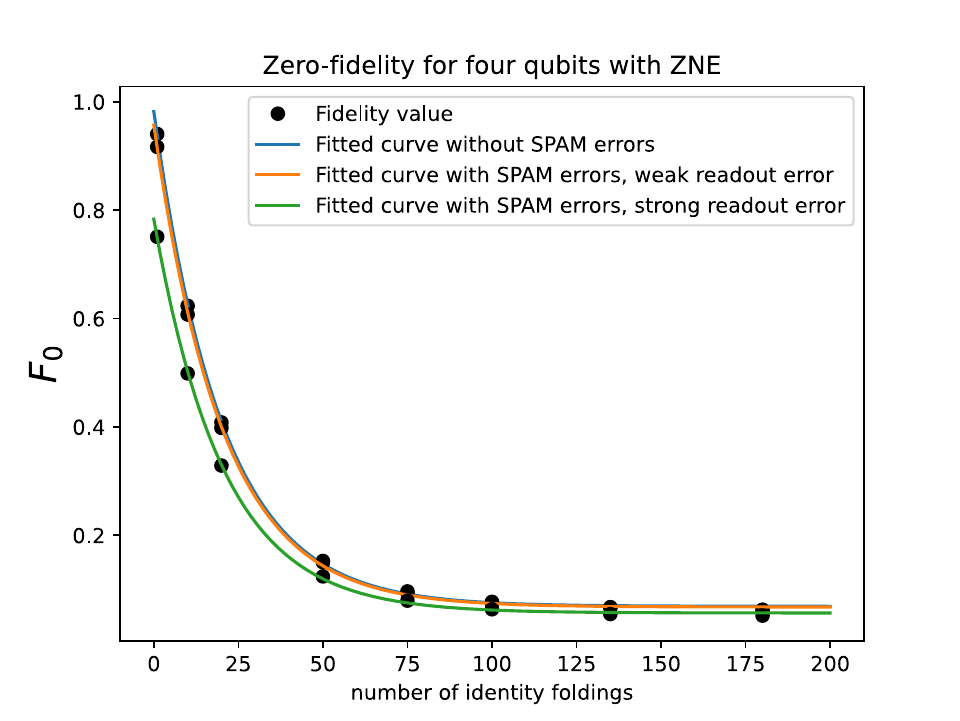}
        \caption{$p \simeq 0.952$ for all curves.}
        \label{Four_zero_channel_RB}
    \end{subfigure}%
    \begin{subfigure}[t]{0.32\textwidth}
        \centering
        \includegraphics[width=\textwidth]{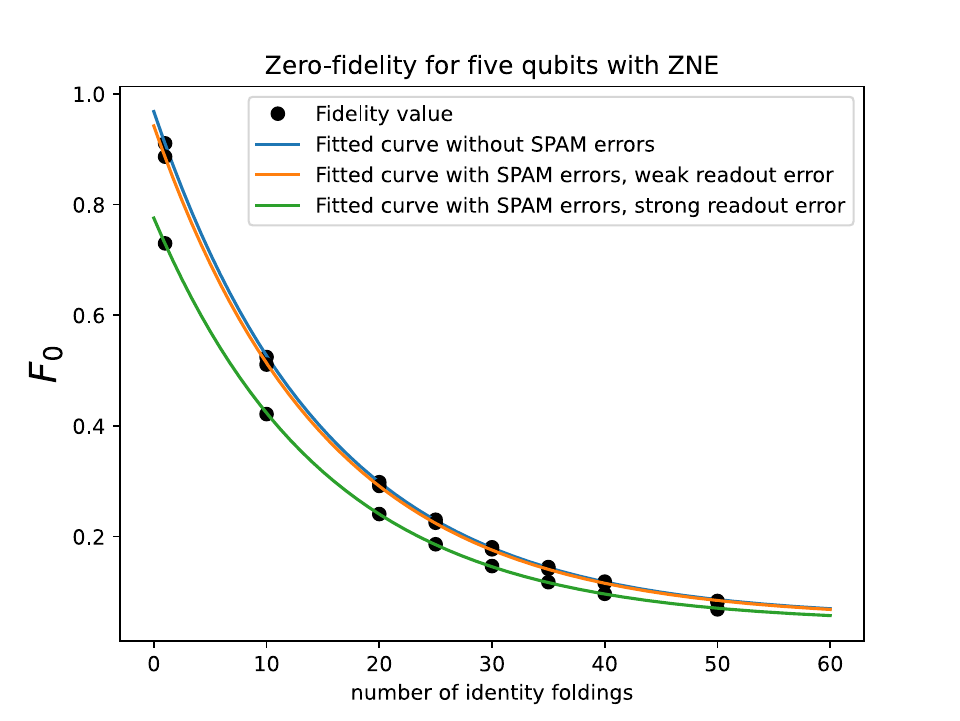}
        \caption{$p \simeq 0.937$ for all curves.}
        \label{Five_zero_channel_RB}
    \end{subfigure}%
    \caption{Zero-fidelity with zero noise extrapolation. }
    \label{fig:zero-fidelity-zne}
\end{figure}

\begin{figure}[htbp]
    \centering
    \captionsetup[subfigure]{position=top, singlelinecheck=false,skip=0pt}%
    \begin{subfigure}[t]{0.42\textwidth} 
        \centering
        \includegraphics[width=0.75\textwidth]{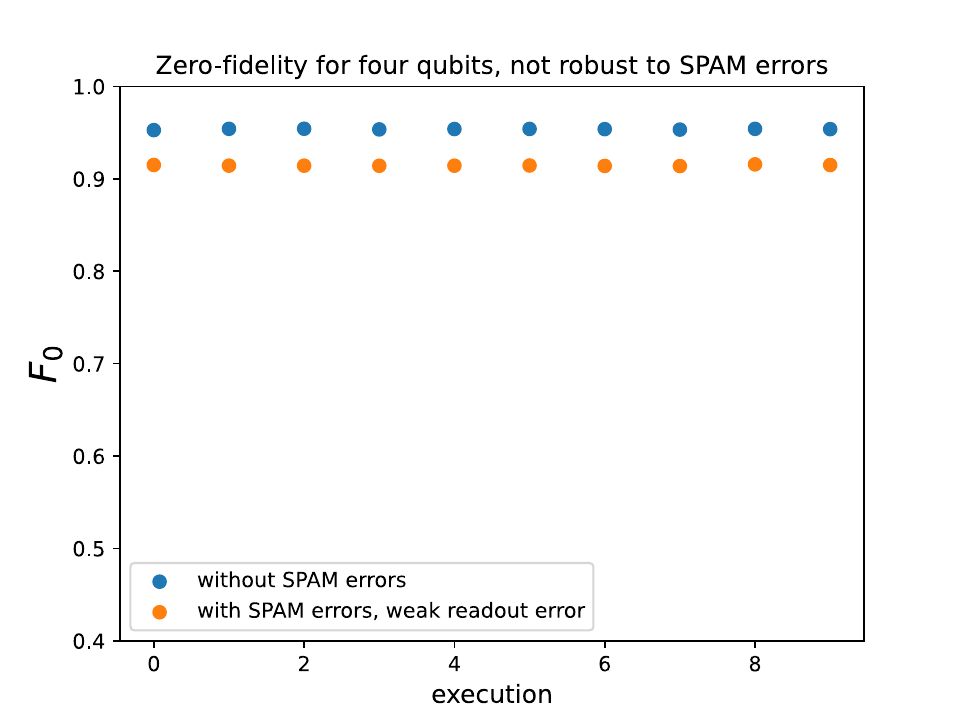} 
        \caption{Avg. zero-fidelity without SPAM errors: $0.953$, avg. zero-fidelity with SPAM errors: $0.914$. }
        \label{Four_qubit_zero_fidelity}
    \end{subfigure}%
    \hfill 
    \begin{subfigure}[t]{0.42\textwidth}
        \centering
        \includegraphics[width=0.75\textwidth]{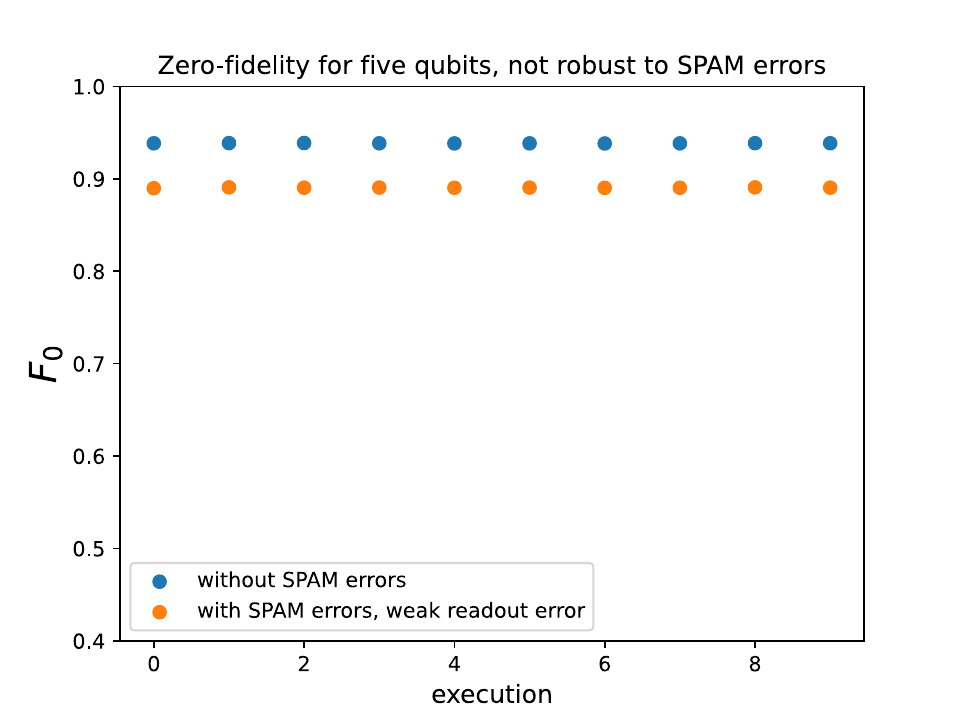}
        \caption{Avg. zero-fidelity without SPAM errors: $0.938$, avg. zero-fidelity with SPAM errors: $0.890$. }
        \label{Five_qubits_zero}
    \end{subfigure}%
    \caption{Zero-fidelity for four and five qubits. Depolarizing error was added on CZ gates for the executions with SPAM errors. $\lambda$ was set to $0.01$.}
    \label{fig:zero-fidelity-zne-2}
\end{figure}
\end{widetext}

Specifically, the extracted decay rates precisely match the ideal, SPAM-free single-circuit fidelities: 
\begin{itemize}
	\item For three qubits (Fig.~\ref{Zero_fidelity_channels}), the decay rate was $p\simeq0.966$ across all curves. This is almost the same as the fidelity without SPAM errors calculated by using single circuit length in Fig.~\ref{fig:CZ_SPAM_fixed_gate_error}
	\item For four qubits (Fig.~\ref{Four_zero_channel_RB}), the decay rate was $p\simeq0.952$, aligning with the SPAM-free baseline of 0.953 (compared to the unmitigated SPAM average of 0.914 in Fig.~\ref{Four_qubit_zero_fidelity}).
	\item For five qubits (Fig.~\ref{Five_zero_channel_RB}), the decay rate was $p\simeq0.937$, aligning with the SPAM-free baseline of 0.938 (compared to the unmitigated SPAM average of 0.890 in Fig.~\ref{Five_qubits_zero}).
\end{itemize}
These results confirm that zero noise extrapolation successfully decouples state preparation and measurement errors from the zero-fidelity estimation.

\section{Conclusion} \label{Sec5}

In this work, we introduced two methodologies to render the zero-fidelity protocol robust against SPAM errors: integration with randomized benchmarking (RB) and the application of zero noise extrapolation (ZNE). Our numerical simulations validate both approaches, confirming that intrinsic channel fidelity can be accurately isolated from SPAM-induced distortions. While combining zero-fidelity with RB successfully mitigates errors for small-scale systems (up to three qubits), the approach is fundamentally bottlenecked by the exponential growth of the Clifford group. Conversely, ZNE bypasses this limitation entirely. By utilizing identity folding rather than random Clifford sampling, ZNE significantly reduces memory requirements and circuit generation overhead. This approach allowed us to efficiently scale our SPAM-independent zero-fidelity estimations to five-qubit systems. Future research will explore the application of these mitigation strategies to error models beyond depolarizing noise channels. Additionally, we plan to investigate other scalable characterization protocols, such as cycle benchmarking \cite{cycle}, to further refine the accuracy and experimental efficiency of SPAM-independent zero-fidelity evaluations.

\begin{acknowledgments}
The authors wish to acknowledge the support of the National Taiwan University's High Performance Computing facilities, on which our numerical results were obtained. Yu-Hao Chen and Renata Wong acknowledge support from the National Science and Technology Council grant No. NSTC 114-2112-M-182-002-MY3. Renata Wong acknowledges support from Chang Gung Memorial Hospital grant No. BMRPL94. H.-S.\ Goan acknowledges support from the National Science and Technology Council (NSTC), Taiwan, under Grants No.\ NSTC 113-2112-M-002-022-MY3, No.\ NSTC 113-2119M-002-021, No.\ NSTC 114-2119-M-002-018, No.\ NSTC 114-2119-M-002-017-MY3, and NSTC 115-2119-M-002-005,  the support of Taiwan Semiconductor Research Institute (TSRI) through the Joint Developed Project (JDP), and from the Physics Division, National Center for Theoretical Sciences (NCTS), Taiwan. H.-S.\ Goan also acknowledges support from the National Taiwan University (NTU) under Grants No.\ NTU-CC-115L8937, No.\ NTU-CC-115L893704 and No.\ NTU-CC-115L8512, as well as support from the Center for Advanced Computing and Imaging in Biomedicine (NTU-115L900702) through the Featured Areas Research Center Program within the framework of the Higher Education Sprout Project by the Ministry of Education (MOE), Taiwan.
\end{acknowledgments}

\section{Code availability}
The code used for this paper can be found at
~\\
\url{https://github.com/plain12356/RB_with_zero_fidelity}

\appendix

\section{Another form of the process fidelity}\label{appendix-a}

We derive the equation \begin{equation*}
    F(\Lambda, \Gamma)  = \frac{1}{4^n}\sum^{{4^n}}_{{i,j = 1}}C_{ij}\mathbf{Tr}[\Gamma(\rho_{i})W_{j}] \end{equation*} with $ C_{ij} = \sum^{{4^n}}_{{l = 1}}[B^{-1}]_{li}\mathbf{Tr}[\Lambda(\rho_l)W_j]$ from the equation \begin{equation*}
     F(\Lambda, \Gamma) = \frac{1}{4^n}\sum^{4^n}_{i = 1}\mathbf{Tr}[\Lambda(\sigma^{\dagger}_{i})\Gamma(\sigma_i)],
\end{equation*} where $W_j$ is a set of Pauli operators, $\rho_i$ are the quantum states which are prepared by the eigenstates of Pauli operators, and $\Lambda, \Gamma$ are the ideal channel and the actual channel, respectively. Before the derivation we introduce some notational conventions.

\subsection{Preliminary notation} \label{A.1}
We define the vectorization of a matrix, which is a linear transformation converting a matrix into a vector form.

Let $A$ be a $m \times n$ matrix and let $\mathbf{Vec}(A)$ denote a vectorization of the matrix $A$ given as \begin{equation*}
    \mathbf{Vec}(A) = [a_{1,1}, ...a_{m,1}, a_{1,2},...,a_{m,2},...,a_{m,n}]^{\mathbf{T}},
\end{equation*} where ${\mathbf{T}}$ stands for transposition. Here is an example: 
\begin{equation*}
    M = \begin{pmatrix}
        1 & 2 \\
3 & 4 \end{pmatrix} \quad \rightarrow \quad \mathbf{Vec}(M) =  \begin{pmatrix}
1 \\ 3 \\
2 \\ 4 
    \end{pmatrix}. 
\end{equation*} 
The vectorization of a product of three matrices $ABC$ is given as \begin{equation} \label{A.6}
    \mathbf{Vec}(ABC) = (C^{\mathbf{T}}\otimes A)\mathbf{Vec}(B)
\end{equation} 
and the inner product (Hilbert–Schmidt inner product) between two vectorized matrices is \begin{equation*}
    \mathbf{Vec}(A)^{\dagger}\mathbf{Vec}(B) = \mathbf{Tr}(A^{\dagger}B).
\end{equation*} 

For convenience we denote $\mathbf{Vec}(A) = |A\rangle\rangle $ and $ \mathbf{Vec}(A)^{\dagger} = \langle\langle A|.$
Now assume a density matrix $\rho$ that is evolved by a channel $\mathcal{U}$, which can be represented as \begin{equation*}
    \rho' \rightarrow \mathcal{U}(\rho) = U\rho U^{\dagger}, 
\end{equation*} where $U$ is a unitary matrix. Because of equation (\ref{A.6}), $\rho'$ can be rewritten as 
\begin{align}
    \mathbf{Vec}(\rho') &=  \mathbf{Vec}(\mathcal{U}(\rho)) \nonumber \\ 
    &= \mathbf{Vec}(U\rho U^{\dagger}) \nonumber \\ 
    &= (U^{\dagger}\otimes U)\mathbf{Vec}(\rho) \nonumber \\ &= \label{A.12}\mathcal{U}|\rho\rangle\rangle .
\end{align} Here we use the superoperator representation to describe the channel $\mathcal{U} = (U^{\dagger}\otimes U).$ It can be understood that the superoperator $\mathcal{U}$ in term (\ref{A.12}) is an operator and the vectorized $\rho$ is a vector.

\subsection{Derivation}
We can now vectorize the Pauli operator \begin{equation*}
    P_i \rightarrow |P_i\rangle\rangle/\sqrt{2^n},
\end{equation*} where $n$ is the number of qubits. For $n = 1$ the vectorized Pauli operators can be represented as \begin{equation*}
    |P_i\rangle\rangle \in \{   |I\rangle\rangle  \quad |X\rangle\rangle \quad |Y\rangle\rangle\quad |Z\rangle\rangle\},
\end{equation*} where \begin{equation*}
    |I\rangle\rangle = \begin{pmatrix}
1 \\ 0 \\
0 \\ 1 \end{pmatrix}  |X\rangle\rangle = \begin{pmatrix}
0 \\ 1 \\
1 \\ 0 \end{pmatrix}  |Y\rangle\rangle = \begin{pmatrix}
0 \\ i \\
-i \\ 0 \end{pmatrix}  |Z\rangle\rangle = \begin{pmatrix}
1 \\ 0 \\
0 \\ -1 \end{pmatrix}.
\end{equation*}

Since the density matrix can be represented by the Pauli operator, we represent a quantum state as \begin{equation} \label{A.14}
    |\rho\rangle\rangle = \sum_i c_i|P_i\rangle\rangle,
\end{equation} where $c_i = \langle\langle P_i|\rho\rangle\rangle$ (we have normalized Pauli operators in order to avoid the factor). Note that the coefficients $c_i$ can be considered to be the expectation values of Pauli operators with respect to the quantum state $\rho$. Putting the coefficients $c_i$ back into Eq.~(\ref{A.14}), the quantum state becomes \begin{equation} \label{A.17}
    |\rho\rangle\rangle = \sum_i |P_i\rangle\rangle\langle\langle P_i|\rho\rangle\rangle .
\end{equation} Consider another quantum state \begin{equation} \label{A.18}
    |\sigma\rangle\rangle = \sum_j |P_j\rangle\rangle\langle\langle P_j|\sigma\rangle\rangle,
\end{equation} one can perform the Hilbert–Schmidt inner product to obtain \begin{align} \label{A.19}
    \mathbf{Tr}[\sigma^{\dagger}\rho] &= \sum_i\sum_j \langle\langle\sigma|[ P_j \rangle\rangle(\langle\langle P_j|P_i\rangle\rangle)\langle\langle P_i]|\rho\rangle\rangle  \\&= \sum_i\langle\langle\sigma|P_i\rangle\rangle\cdot\langle\langle P_i|\rho\rangle\rangle  \\ &=   \label{A.21}\sum_{i = 1}\mathbf{Tr}[\sigma P_i]\mathbf{Tr}[\rho P_i].
\end{align} Since Pauli operators are mutually orthogonal, this means that	
$$\begin{cases}
\mathbf{Tr}[P_iP_j] = 0, & \mbox{if } i \ne j \\
\mathbf{Tr}[P_iP_j] = 1, & \mbox{if } i = j
\end{cases}$$
and recall that $\langle\langle P_i|\rho\rangle\rangle = \mathbf{Tr}[\rho P_i].$ Now, back to the equation \begin{equation*}
    F(\Lambda, \Gamma) = \frac{1}{4^2}\sum^{4^2}_{i = 1}\mathbf{Tr}[\Lambda(\sigma^{\dagger}_{i})\Gamma(\sigma_i)],
\end{equation*}
where $\{\sigma_i\}^{4}_{i = 1}$ is a set of Pauli operators, for $n$ qubits $\sigma_i \in \{I, X, Y, Z\}^{\otimes n}.$

We can rewrite \begin{equation*}
    F(\Lambda, \Gamma) = \frac{1}{4^n}\sum^{4^n}_{i = 1}\mathbf{Tr}[\Lambda(\sigma^{\dagger}_i)\Gamma(\sigma_i)] 
\end{equation*} as
\begin{equation} \label{A.24}
    F(\Lambda, \Gamma) = \frac{1}{4^n}\sum^{4^n}_{i,j = 1}[B^{-1}]_{ij}\mathbf{Tr}[\Lambda(\rho_i)\Gamma(\rho_j)],
\end{equation}
where $B_{ij} = \mathbf{Tr}[\rho^{\dagger}_i\rho_j]$ and $\rho_i,\rho_j$ are density matrices that can be prepared from the eigenstates of Pauli operators. Let $\rho' = \Lambda(\rho)$ and $\sigma = \Gamma(\rho)$ so Eq.~(\ref{A.24}) becomes \begin{equation}
    F(\Lambda, \Gamma) = \frac{1}{4^n}\sum^{4^n}_{i,j = 1}[B^{-1}]_{ij}\mathbf{Tr}[\rho'_i\sigma_j].
\end{equation} From Eqs.~(\ref{A.19}) through (\ref{A.21}) we  obtain 
\begin{align*}
     F(\Lambda, \Gamma) &= \frac{1}{4^n}\sum^{4^n}_{i,j = 1}[B^{-1}]_{ij}\mathbf{Tr}[\rho'_i\sigma_j] \\ &= \frac{1}{4^n}\sum^{4^n}_{i,j = 1}\sum^{4^n}_{k =1}[B^{-1}]_{ij}\mathbf{Tr}[\sigma_j P_k]\mathbf{Tr}[\rho'_i P_k] \\ &= \frac{1}{4^n}\sum^{4^n}_{i,j = 1}\sum^{4^n}_{k =1}[B^{-1}]_{ij}\mathbf{Tr}[\Gamma(\rho_j) P_k]\mathbf{Tr}[\Lambda(\rho_i) P_k].
\end{align*}

 As mentioned in the beginning of this appendix, we finally get 
\begin{equation*}
    F(\Lambda, \Gamma)  = \frac{1}{4^n}\sum^{{4^n}}_{{i,j = 1}}C_{ij}\mathbf{Tr}[\Gamma(\rho_{i})W_{j}]
\end{equation*} with $C_{ij} = \sum^{{4^n}}_{{l = 1}}[B^{-1}]_{li}\mathbf{Tr}[\Lambda(\rho_l)W_j],$ where $W_j$ is a set of Pauli operators.

\section{Decay rate of the zero-fidelity}

We show that the equation \begin{equation*}
    F^m(\mathcal{U}, \tilde{\mathcal{U}}) = \frac{1}{4^n}\sum^{4^n}_{i, j =1}\mathbf{Tr}[\mathcal{U}^m(\rho_i)W_j]\mathbf{Tr}[\tilde{\mathcal{U}}^m(\rho_i) W_j],
\end{equation*} where $\rho_i$ are the quantum state prepared by the SIC-states and $W_j$ is a set of Pauli operators, is equal to  \begin{equation} \label{B.31}
    A_0p^m+B_0,
\end{equation} where the coefficients $A_0$ and $B_0$ absorb state preparation and measurement errors, and $m$ is the number of Clifford operators.

Since the process fidelity \begin{equation*}
    F(\Lambda, \Gamma) = \frac{1}{4^n}\sum^{4^n}_{i = 1}\mathbf{Tr}[\Lambda(\sigma^{\dagger}_{i})\Gamma(\sigma_i)]
\end{equation*} can be written as follows (this was established in Appendix A) \begin{equation*}
    F(\Lambda, \Gamma) = \frac{1}{4^n}\sum^{4^n}_{i = 1}\mathbf{Tr}[\Lambda(\rho_i)W_j]\mathbf{Tr}[\Gamma(\rho_i)W_j],
\end{equation*} we only need to prove that the process fidelity with the channel applied $m$ times, i.e.,  \begin{equation*}
    F^m(\Lambda, \Gamma) = \frac{1}{4^n}\sum^{4^n}_{i = 1}\mathbf{Tr}[\Lambda^m(\sigma^{\dagger}_{i})\Gamma^m(\sigma_i)]
\end{equation*} is equal to Eq.~(\ref{B.31}).

    As shown in Section \ref{Sec3}, \begin{align*}
    F^m(\mathcal{U}, \tilde{\mathcal{U}}) &= \frac{1}{4^n}\sum^{4^n}_{i = 1}\mathbf{Tr}[\mathcal{U}^m(\rho_i)\tilde{\mathcal{U}}^m(\rho_i)] \\ & = A_0p^m + B_0.
\end{align*} We have also shown that \begin{align*}
    F(\Lambda, \Gamma) &= \frac{1}{4^n}\sum^{4^n}_{i = 1}\mathbf{Tr}[\Lambda(\sigma^{\dagger}_i)\Gamma(\sigma_i)] \\ &= \frac{1}{4^n}\sum^{4^n}_{i, j = 1}C_{ij}\mathbf{Tr}[\Lambda(\rho_i)W_j], 
\end{align*} where \begin{equation}
    C_{ij} = \sum^{4^n}_{l=1}[B^{-1}]_{li}\mathbf{Tr}[\Gamma(\rho_i)W_j].
\end{equation} Furthermore, if the density matrix $\rho_i$ is prepared by the eigenstates of Pauli operators, we can write the process fidelity after \cite{cycle} as \begin{equation} \label{B.40}
    F(\Lambda, \Gamma) = \frac{1}{4^n}\sum^{4^n}_{i,j}2^{-n}\mathbf{Tr}[\Gamma(\rho_i)W_j]\mathbf{Tr}[\Lambda(\rho_i)W_j].
\end{equation} Now let the quantum states be prepared by the SIC-states instead of the eigenstates of Pauli operators. In this case, the coefficient $2^{-n}$ in Eq.~(\ref{B.40}) is $1$ and then Eq.(\ref{B.40}) becomes \begin{equation*}
    F(\Lambda, \Gamma) = \frac{1}{4^n}\sum^{4^n}_{i,j}\mathbf{Tr}[\Gamma(\rho_i)W_j]\mathbf{Tr}[\Lambda(\rho_i)W_j].
\end{equation*} Finally, it can be concluded that \begin{align*}
    F^m(\Lambda, \Gamma) &= \frac{1}{4^n}\sum^{4^n}_{i,j}\mathbf{Tr}[\Gamma^m(\rho_i)W_j]\mathbf{Tr}[\Lambda^m(\rho_i)W_j] \\ &= A_0p^m+B_0.
\end{align*}

\section{The physical meaning of parameter $p$ in channel noise scaling} \label{Appen C}

In this appendix, we analyze the physical meaning of parameter $p$. This parameter can be obtained by fitting the results of the zero-fidelity. The decrease in the fidelity is a result of applying quantum circuits repeatedly, i.e., letting channel noise to accumulate. We refer to this method as \textit{identity folding}. First, we introduce complex projective designs.

\subsubsection{Complex projective t-design}
The definition of a complex projective design is similar to the definition of a spherical design. Consider a finite subset of unit vectors $\mathbf{X}$ on a sphere $\mathbf{S}(\mathbb{R}^{d})$, where $\mathbb{R}^{d}$ is $d$-dimensional Euclidean space. The following is the definition of a spherical $t$-design based on \cite{spherical_designs,roy2011complexsphericaldesignscodes}.

\newtheorem{definition}{Definition}[section]

\begin{definition}
    Let $t$ be a natural number. A finite subset $\mathbf{X} \subseteq \mathbf{S}(\mathbb{R}^{d})$ is called a spherical t-design if \begin{equation} \label{4.41}
    \frac{1}{|\mathbf{S}(\mathbb{R}^{d})|}\int_{\mathbf{X} \in \mathbf{S}(\mathbb{R}^{d})} f(\mathbf{X})d\sigma(\mathbf{X}) = \frac{1}{|\mathbf{X}|} \sum_{\mathbf{u} \in \mathbf{X}}
    f(\mathbf{u})
\end{equation} holds for any polynomial $f(\mathbf{X}) =f(x_1, x_2, . . . , x_n)$ of degree at most $t$, with the usual integral on the unit sphere.
\end{definition}

For complex projective $t$-design, instead of on the real $d$-dimensional space $\mathbb{R}^{d}$, the subset of unit vectors $\mathbf{X}$ is in $\mathbf{S}(\mathbb{C}^{d})$. Furthermore the degree of the polynomial function is $t$, meaning the function has degree at most $t$ for all real numbers in the entries of the vectors and at most $t$ for the complex conjugates of these entries. The definition of complex projective $t$-design is given below and follows \cite{Matteo2014ASI}.
\begin{definition}  
Let $p_{t,t}$ be a polynomial that has homogeneous degree $t$ in $d$ variables, and degree $t$ in the complex conjugates of these variables. It can be written as \begin{equation} \label{Definition 4.5.2}
    \frac{1}{|\mathbf{X}|}\sum_{x \in \mathbf{X}}p_{t,t}(x) = \int_{\mathbf{S}(\mathbb{C}^{d})} p_{t,t}(u)d\eta(u).
\end{equation}
\end{definition} From Def.~(\ref{Definition 4.5.2}) we can see that complex projective designs can be written in terms of quantum states. The reason why complex projective designs can be written in such a way is that quantum states are vectors in a Hilbert space (complex space). The following is the form of complex projective designs in terms of quantum states: \begin{definition} \label{Definition 4.5.3}
    Let $\{p_1, p_2,...,p_K \} $ be a probability distribution over quantum states $\{ |\psi_1\rangle, |\psi_2\rangle, \dots, |\psi_K\rangle  \}$. Such a distribution is called a quantum state design if \begin{equation*}
        \sum_{i}p_i(|\psi_t\rangle \langle \psi_t|)^{\otimes t} = \int_{\Psi}d\Psi (|\Psi\rangle \langle \Psi|)^{\otimes t}.
    \end{equation*}
\end{definition} Def.~\ref{Definition 4.5.3} means that quantum state design is a probability distribution over a finite set of quantum states \cite{Matteo2014ASI}. From Ref.~\cite{Q} we know that a unitary $t$-design induces quantum state $t$-designs for the probability distribution over states $U|\psi_0\rangle$, where $U$ is chosen from unitary group with fixed state $|\psi_0\rangle$ \cite{Haferkamp2022randomquantum}. The relation between unitary designs and quantum state designs is important to our analysis. The reason is that the four SIC-states are the quantum state $2$-design \cite{Matteo2014ASI, 1523643} which means that we can make a connection between SIC-states and the unitary $2$-design. Therefore, we can sum up the SIC-states with a fixed unitary $U$. 

\subsubsection{Quantum state design and depolarizing channel}
So far we have introduced spherical $t$-designs, and complex projective $t$-designs also known as quantum state $t$-designs. We also mentioned the relationship between quantum state $t$-designs and unitary $t$-designs. 
Since the four SIC-states are quantum state $2$-design, we would like to justify that the parameter $p$ obtained by fitting the results of the zero-fidelity (where the initial states are SIC-states) with the method of repeatedly applying circuits (identity folding) has a physical meaning. 

Here, we show that the physical meaning of parameter $p$ obtained by fitting the results of the zero-fidelity corresponds to the strength of depolarizing channel. Therefore, with the decay rate we can learn the channel fidelity between an ideal channel and an actual channel without considering the effects from SPAM errors.

We now consider the formula for average gate fidelity \begin{equation} \label{4.44}
    \mathbb{E}(F_g) = \int_{U(D)}dU \mathbf{Tr}[\rho U^{-1}\Lambda_{\mathbf{c}}(U\rho U^{-1})U], 
\end{equation} where $dU$ is the unitarily invariant Haar measure, $\rho$ is a quantum state $|\psi\rangle\langle\psi|$ and $\Lambda_{\mathbf{c}}$ is a noisy channel. Eq.~(\ref{4.44}) means that the gate fidelity is averaged over the unitary group. 
As shown in Ref. \cite{EmersonRBCYCLE}, Eq.~(\ref{4.44}) can be written as \begin{equation*}
    \mathbb{E}_{U}(F_g) = \mathbf{Tr}[\rho\hat{\Lambda}^{\mathbf{avg}}_{\mathbf{c}}\rho] = F_g(\hat{\Lambda}^{\mathbf{avg}}_{\mathbf{c}}), 
\end{equation*} where $\hat{\Lambda}^{\mathbf{avg}}_{\mathbf{c}} \equiv \int dU\hat{U}\hat{\Lambda_{\mathbf{c}}}\hat{U}^{-1}.$ Based on the analysis of $\hat{\Lambda}^{\mathbf{avg}}_{\mathbf{c}}$ in Ref. \cite{EmersonRBCYCLE}, the noisy channel $\hat{\Lambda}^{\mathbf{avg}}_{\mathbf{c}}$ can be written as follows \begin{equation} \label{4.46}
    \hat{\Lambda}^{\mathbf{avg}}_{\mathbf{c}}\rho = p\rho + (1-p)\frac{\mathbb{I}}{D},
\end{equation}  where $p$ can be seen as the strength of the depolarizing channel. By applying the channel $m$ times, we can obtain the fidelity in the form \begin{equation*}
    F^m = p^m\mathbf{Tr}[\rho(0)^2] + \frac{(1-p^m)}{D}. 
\end{equation*} If $p = 1 $ then the fidelity $F^m$ is equal to $1$, which means there is no error. It is worth noting that the mathematical form of the depolarizing channel in Ref.~\cite{EmersonRBCYCLE} is different from the mathematical form in Ref.~\cite{PhysRevA.108.032602}, where it is written as \begin{equation} \label{4.48}
    \Lambda^{d}_{{\mathbf{c}}, p}(\rho) = (1-p)\rho+p\mathbf{Tr}(\rho)\frac{I_d}{d}, 
\end{equation} where $d$ is the dimension of the Hilbert space and $p$ is the depolarizing probability. Therefore, the physical interpretations of these two mathematical forms may be a little different. So far we have mentioned the relation between the average gate fidelity and the depolarizing channel; and the physical meaning $p$ in the depolarizing channel written as Eqs.~(\ref{4.46}) and (\ref{4.48}), respectively. Now we can use the equivalence mentioned in Ref.~\cite{EmersonRBCYCLE} \begin{equation}\label{4.49}
    \mathbb{E}_{\psi}(F_g) =  \mathbb{E}_{U}(F_g),
\end{equation} where \begin{equation}\label{4.50}
    \mathbb{E}_{\psi}(F_g) \equiv \int d\psi\langle\psi|U^{-1}(\Lambda(U|\psi\rangle\langle\psi|U^{-1}))U|\psi\rangle.
\end{equation} The measure $d\psi$ in Eq.~(\ref{4.50}) is unitarily invariant over the set of pure states, and $|\psi\rangle$ is a quantum state. 

Finally, we can fit our results of the zero-fidelity to exponential decay curve $Ap^{m} +B$, where $p$ is the strength of depolarization and $m$ is the number of repeated applications of circuits (identity folding), and $A, B$ are caused by SPAM errors as long as our noisy channel is the depolarizing channel \cite{GiurgicaTiron2020DigitalZN}. So, for the ideal situation (without depolarizing noise and SPAM errors), the fidelity is $p = 1$. Furthermore, based on the analysis in Ref.~\cite{EmersonRBCYCLE}, we can explain the relation between quantum state $2$-design (recall that SIC-state are quantum state $2$-design) and unitary $2$-design. So we can twirl the noise channel into depolarizing channel \cite{EmersonRBCYCLE}. We can conclude that zero noise extrapolation can be used to obtain the channel fidelity from the decay rate of the zero-fidelity without considering the effects from SPAM errors.


\bibliography{apssamp}

@article{Flammia,
  title = {Direct Fidelity Estimation from Few Pauli Measurements},
  author = {Flammia, Steven T. and Liu, Yi-Kai},
  journal = {Phys. Rev. Lett.},
  volume = {106},
  issue = {23},
  pages = {230501},
  numpages = {4},
  year = {2011},
  month = {Jun},
  publisher = {American Physical Society},
  doi = {10.1103/PhysRevLett.106.230501},
  url = {https://link.aps.org/doi/10.1103/PhysRevLett.106.230501}
}

@article{Zero,
  title = {Efficient assessment of process fidelity},
  author = {Greenaway, Sean and Sauvage, Fr\'ed\'eric and Khosla, Kiran E. and Mintert, Florian},
  journal = {Phys. Rev. Res.},
  volume = {3},
  issue = {3},
  pages = {033031},
  numpages = {15},
  year = {2021},
  month = {Jul},
  publisher = {American Physical Society},
  doi = {10.1103/PhysRevResearch.3.033031},
  url = {https://link.aps.org/doi/10.1103/PhysRevResearch.3.033031}
}

@article{randomized3,
  title = {Randomized benchmarking of quantum gates},
  author = {Knill, E. and Leibfried, D. and Reichle, R. and Britton, J. and Blakestad, R. B. and Jost, J. D. and Langer, C. and Ozeri, R. and Seidelin, S. and Wineland, D. J.},
  journal = {Phys. Rev. A},
  volume = {77},
  issue = {1},
  pages = {012307},
  numpages = {7},
  year = {2008},
  month = {Jan},
  publisher = {American Physical Society},
  doi = {10.1103/PhysRevA.77.012307},
  url = {https://link.aps.org/doi/10.1103/PhysRevA.77.012307}
}

@article{randomized4,
author = {Erika Kawakami  and Thibaut Jullien  and Pasquale Scarlino  and Daniel R. Ward  and Donald E. Savage  and Max G. Lagally  and Viatcheslav V. Dobrovitski  and Mark Friesen  and Susan N. Coppersmith  and Mark A. Eriksson  and Lieven M. K. Vandersypen },
title = {Gate fidelity and coherence of an electron spin in an Si/SiGe quantum dot with micromagnet},
journal = {Proceedings of the National Academy of Sciences},
volume = {113},
number = {42},
pages = {11738-11743},
year = {2016},
doi = {10.1073/pnas.1603251113},
URL = {https://www.pnas.org/doi/abs/10.1073/pnas.1603251113},
eprint = {https://www.pnas.org/doi/pdf/10.1073/pnas.1603251113}}

@article{programmable,
	doi = {10.1038/nature25766},
  
	url = {https://doi.org/10.1038%2Fnature25766},
  
	year = 2018,
	month = {feb},
  
	publisher = {Springer Science and Business Media {LLC}
},
  
	volume = {555},
  
	number = {7698},
  
	pages = {633--637},
  
	author = {T. F. Watson and S. G. J. Philips and E. Kawakami and D. R. Ward and P. Scarlino and M. Veldhorst and D. E. Savage and M. G. Lagally and Mark Friesen and S. N. Coppersmith and M. A. Eriksson and L. M. K. Vandersypen},
  
	title = {A programmable two-qubit quantum processor in silicon},
  
	journal = {Nature}
}

@article{SIC,
    author = {Renes, Joseph M. and Blume-Kohout, Robin and Scott, A. J. and Caves, Carlton M.},
    title = "{Symmetric informationally complete quantum measurements}",
    journal = {Journal of Mathematical Physics},
    volume = {45},
    number = {6},
    pages = {2171-2180},
    year = {2004},
    month = {05},
    issn = {0022-2488},
    doi = {10.1063/1.1737053},
    url = {https://doi.org/10.1063/1.1737053},
    eprint = {https://pubs.aip.org/aip/jmp/article-pdf/45/6/2171/8173125/2171\_1\_online.pdf},
}

@article{cycle,
	doi = {https://doi.org/10.1038/s41467-019-13068-7},
  
	url = {https://doi.org/10.1038%2Fs41467-019-13068-7},
  
	year = 2019,
	month = {nov},
  
	publisher = {Springer Science and Business Media {LLC}
},
  
	volume = {10},
  
	number = {1},
  
	author = {Alexander Erhard and Joel J. Wallman and Lukas Postler and Michael Meth and Roman Stricker and Esteban A. Martinez and Philipp Schindler and Thomas Monz and Joseph Emerson and Rainer Blatt},
  
	title = {Characterizing large-scale quantum computers via cycle benchmarking},
  
	journal = {Nature Communications}
}

@article{EmersonRBCYCLE,
doi = {10.1088/1464-4266/7/10/021},
url = {https://dx.doi.org/10.1088/1464-4266/7/10/021},
year = {2005},
month = {sep},
publisher = {},
volume = {7},
number = {10},
pages = {S347},
author = {Joseph Emerson and Robert Alicki and Karol Życzkowski},
title = {Scalable noise estimation with random unitary operators},
journal = {Journal of Optics B: Quantum and Semiclassical Optics}
}

@article{AVE,
title = {Fidelity of single qubit maps},
journal = {Physics Letters A},
volume = {294},
number = {5},
pages = {258-260},
year = {2002},
issn = {0375-9601},
doi = {https://doi.org/10.1016/S0375-9601(02)00069-5},
url = {https://www.sciencedirect.com/science/article/pii/S0375960102000695},
author = {Mark D. Bowdrey and Daniel K.L. Oi and Anthony J. Short and Konrad Banaszek and Jonathan A. Jones}
}

@article{RBCIRCUIT,
  title = {Improved simulation of stabilizer circuits},
  author = {Aaronson, Scott and Gottesman, Daniel},
  journal = {Phys. Rev. A},
  volume = {70},
  issue = {5},
  pages = {052328},
  numpages = {14},
  year = {2004},
  month = {Nov},
  publisher = {American Physical Society},
  doi = {10.1103/PhysRevA.70.052328},
  url = {https://link.aps.org/doi/10.1103/PhysRevA.70.052328}
}

@misc{Qiskit,
    author = {{Qiskit contributors}},
    title = {Qiskit: An Open-source Framework for Quantum Computing},
    year = {2023},
    doi = {10.5281/zenodo.2573505}
}

@article{Threerb,
  title = {Three-Qubit Randomized Benchmarking},
  author = {McKay, David C. and Sheldon, Sarah and Smolin, John A. and Chow, Jerry M. and Gambetta, Jay M.},
  journal = {Phys. Rev. Lett.},
  volume = {122},
  issue = {20},
  pages = {200502},
  numpages = {6},
  year = {2019},
  month = {May},
  publisher = {American Physical Society},
  doi = {10.1103/PhysRevLett.122.200502},
  url = {https://link.aps.org/doi/10.1103/PhysRevLett.122.200502}
}

@article{RBassumption,
  title = {Characterizing quantum gates via randomized benchmarking},
  author = {Magesan, Easwar and Gambetta, Jay M. and Emerson, Joseph},
  journal = {Phys. Rev. A},
  volume = {85},
  issue = {4},
  pages = {042311},
  numpages = {16},
  year = {2012},
  month = {Apr},
  publisher = {American Physical Society},
  doi = {10.1103/PhysRevA.85.042311},
  url = {https://link.aps.org/doi/10.1103/PhysRevA.85.042311}
}

@article{IRB,
  title = {Efficient Measurement of Quantum Gate Error by Interleaved Randomized Benchmarking},
  author = {Magesan, Easwar and Gambetta, Jay M. and Johnson, B. R. and Ryan, Colm A. and Chow, Jerry M. and Merkel, Seth T. and da Silva, Marcus P. and Keefe, George A. and Rothwell, Mary B. and Ohki, Thomas A. and Ketchen, Mark B. and Steffen, M.},
  journal = {Phys. Rev. Lett.},
  volume = {109},
  issue = {8},
  pages = {080505},
  numpages = {5},
  year = {2012},
  month = {Aug},
  publisher = {American Physical Society},
  doi = {10.1103/PhysRevLett.109.080505},
  url = {https://link.aps.org/doi/10.1103/PhysRevLett.109.080505}
}

@article{mitigation1,
  title = {Practical Quantum Error Mitigation for Near-Future Applications},
  author = {Endo, S. and Benjamin, S. C. and Li, Y.},
  journal = {Phys. Rev. X},
  volume = {8},
  issue = {3},
  pages = {031027},
  numpages = {21},
  year = {2018},
  publisher = {American Physical Society},
  doi = {10.1103/PhysRevX.8.031027},
  url = {https://link.aps.org/doi/10.1103/PhysRevX.8.031027}
}

@article{mitigation2,
  title = {Error Mitigation for Short-Depth Quantum Circuits},
  author = {Temme, K. and Bravyi, S. and Gambetta, J. M.},
  journal = {Phys. Rev. Lett.},
  volume = {119},
  issue = {18},
  pages = {180509},
  numpages = {5},
  year = {2017},
  publisher = {American Physical Society},
  doi = {10.1103/PhysRevLett.119.180509},
  url = {https://link.aps.org/doi/10.1103/PhysRevLett.119.180509}
}

@misc{Matteo2014ASI,
  title={A short introduction to unitary 2-designs},
  author={O. {Di Matteo}},
  year={2014},
  url={https://api.semanticscholar.org/CorpusID:52059910},
    howpublished={unpublished, https://api.semanticscholar.org/CorpusID:52059910, (accessed:{ June 2024})}
}

@INPROCEEDINGS{1523643,
  author={Klappenecker, A. and Rotteler, M.},
  booktitle={Proceedings. International Symposium on Information Theory, 2005. ISIT 2005.}, 
  title={Mutually unbiased bases are complex projective 2-designs}, 
  year={2005},
  volume={},
  number={},
  pages={1740-1744},
  doi={10.1109/ISIT.2005.1523643}}

@article{GiurgicaTiron2020DigitalZN,
  title={Digital zero noise extrapolation for quantum error mitigation},
  author={Giurgica-Tiron, Tudor and Hindy, Yousef and LaRose, Ryan and Mari, Andrea and Zeng, William J.},
  journal={2020 IEEE International Conference on Quantum Computing and Engineering (QCE)},
  year={2020},
  pages={306-316},
  url={https://api.semanticscholar.org/CorpusID:218862807}
}

@article{spherical_designs,
title = {A survey on spherical designs and algebraic combinatorics on spheres},
journal = {Eur. J. Comb.},
volume = {30},
number = {6},
pages = {1392-1425},
year = {2009},
issn = {0195-6698},
doi = {https://doi.org/10.1016/j.ejc.2008.11.007},
url = {https://www.sciencedirect.com/science/article/pii/S0195669808002400},
author = {E. Bannai and E. Bannai}
}

@article{roy2011complexsphericaldesignscodes,
  title={Complex Spherical Designs and Codes},
  author={A. Roy and S. Suda},
  journal={J Comb Des},
issue = {3},
  year={2013},
  volume={22},
  url={https://api.semanticscholar.org/CorpusID:118558019},
  pages={105-148}
}

@article{Q,
   title={Introduction to Haar Measure Tools in Quantum Information: A Beginner's Tutorial},
   volume={8},
   ISSN={2521-327X},
   url={http://dx.doi.org/10.22331/q-2024-05-08-1340},
   DOI={10.22331/q-2024-05-08-1340},
   journal={Quantum},
   publisher={Verein zur Forderung des Open Access Publizierens in den Quantenwissenschaften},
   author={Mele, A. A.},
   year={2024},
   pages={1340} }

@article{Haferkamp2022randomquantum,
  doi = {10.22331/q-2022-09-08-795},
  url = {https://doi.org/10.22331/q-2022-09-08-795},
  title = {Random quantum circuits are approximate unitary {$t$}-designs in depth {$O\left(nt^{5+o(1)}\right)$}},
  author = {Haferkamp, Jonas},
  journal = {{Quantum}},
  issn = {2521-327X},
  publisher = {{Verein zur F{\"{o}}rderung des Open Access Publizierens in den Quantenwissenschaften}},
  volume = {6},
  pages = {795},
  month = sep,
  year = {2022}
}

@article{PhysRevA.108.032602,
  title = {Superadditivity effects of quantum capacity decrease with the dimension for qudit depolarizing channels},
  author = {Etxezarreta Martinez, J. and deMarti iOlius, A. and Crespo, P. M.},
  journal = {Phys. Rev. A},
  volume = {108},
  issue = {3},
  pages = {032602},
  numpages = {9},
  year = {2023},
  publisher = {American Physical Society},
  doi = {10.1103/PhysRevA.108.032602},
  url = {https://link.aps.org/doi/10.1103/PhysRevA.108.032602}
}

@article {24759412,
	Title = {Superconducting quantum circuits at the surface code threshold for fault tolerance},
	Author = {Barends, R and Kelly, J and Megrant, A and Veitia, A and Sank, D and Jeffrey, E and White, TC and Mutus, J and Fowler, AG and Campbell, B and Chen, Y and Chen, Z and Chiaro, B and Dunsworth, A and Neill, C and O'Malley, P and Roushan, P and Vainsencher, A and Wenner, J and Korotkov, AN and Cleland, AN and Martinis, John M},
	DOI = {10.1038/nature13171},
	Number = {7497},
	Volume = {508},
	Month = {April},
	Year = {2014},
	Journal = {Nature},
	ISSN = {0028-0836},
	Pages = {500—503},
		URL = {http://arxiv.org/pdf/1402.4848},
}

\end{document}